\documentclass[12pt]{article}
\usepackage{geometry}
\usepackage{a4}
\usepackage{graphicx,subfigure}
\usepackage{epsf}
\usepackage{amsmath}
\usepackage[dvipsnames]{xcolor}
\usepackage{enumitem}
\usepackage{amssymb}
\usepackage{cite}
\usepackage[normalem]{ulem}

\begin{document}
%%%%%%%%%%%%%%%%%%%%%%%%%%%%
\begin{titlepage}
\title{Aspects of entanglement in non-local field theories with fractional Laplacian }
\author{}
\date{% authors are dated
 Pratim Roy
\thanks{\noindent E-mail:~ pratimroy@hri.res.in, pratroy@gmail.com}
\vskip0.4cm
{\sl Harish-Chandra Research Institute, Homi Bhabha National Institute (HBNI), \\
Chhatnag Road, Jhunsi, Allahabad, India 211019}}
\maketitle\abstract{
\noindent
In recent years, various aspects of theoretical models with long range interactions have attracted attention, ranging from out-of-time-ordered correlators to entanglement. In the present paper, entanglement properties of a simple non-local model with long-range interactions in the form of a fractional Laplacian is investigated in both static and a quantum quench scenario. Logarithmic negativity, which is a measure for entanglement in mixed states is calculated numerically. In the static case, it is shown that the presence of long-range interaction ensures that logarithmic negativity decays much slower with distance compared to short-range models. For a sudden quantum quench, the temporal evolution of the logarithmic negativity reveals that, in contrast to short-range models, logarithmic negativity exhibits no revivals for long-range interactions for the time intervals considered. To further support this result, a simpler measure of entanglement, namely the entanglement entropy is also studied for this class of models. }
\end{titlepage}

\section{Introduction}

The physics of long range systems in general, and entanglement in particular have recently been active topics of research. This is primarily because long range models often exhibit novel physical characteristics such as power law fall-off in correlations \cite{Hastings:2005pr}, rather than exponential. This has significant implications for the propagation of correlations throughout the system, which affects properties such as different entanglement measures. In the present paper, a mixed state entanglement measure known as logarithmic negativity along with entanglement entropy is calculated numerically for a class of simple non-local theories with a fractional Laplacian. 

It may be noted that the majority of recent research on entanglement concentrates on entanglement entropy, which is a valid measure of correlation between a system and its complement in pure state. However, it is not a suitable measure of entanglement for mixed states (a simple example being a finite temperature state in free scalar field theory). For instance, entanglement entropy in thermal states receives both classical and quantum contributions and is thus not effective at capturing correlations in these situations. Naturally, the problem of quantifying entanglement in general mixed states has received much attention, and several quantities such as the entanglement of purification, entanglement of formation, entanglement of distillation and entanglement negativity etc. were put forward (an overview may be found in \cite{Horodecki}). The main criterion for an entanglement measure to be a valid one depends on whether it is a monotone under local quantum operations and classical correlations (LOCC). Mathematically, a quantity $E$ that is a faithful measure of entanglement must have the following property,
 \begin{equation}
 \label{LOCC}
 E(\rho) \geq \sum_{i} p_i E(\rho_i),
 \end{equation} 
which states that the entanglement of a state $\rho$ must (on average) be greater than the entanglement of the total entanglement of states $\rho_i$, which are obtained with probability $p_i$ when a LOCC operation is performed on $\rho$. A quantity which satisfies the above condition is said to be a monotone under LOCC. It was proved in \cite{Plenionegativity} that entanglement negativity satisfies condition (\ref{LOCC}). Recently, logarithmic negativity has been investigated in various physical systems \cite{Audenaert:2002xfl, Angel-Ramelli:2020wfo, MohammadiMozaffar:2017chk, Marcovitch:2008sxc} and has also been calculated for scalar and conformal field theories in various scenarios (including quantum quench) in \cite{Calabrese:2012ew, Calabrese:2012nk, Calabrese:2014yza}.

A crucial point investigated in this paper is the effect of long range interactions on entanglement. It is known in the literature \cite{Shiba:2013jja, Basa:2019ywr, Lu:2019xwg} that long-range interactions lead to substantial changes in entanglement entropy, namely, the behaviour of entanglement entropy changes from area law to a volume law. It is therefore interesting to investigate what effect inclusion of long-range interactions has on the logarithmic negativity. A candidate for a physical system which has long-range interactions is a scalar field theory with a fractional Laplacian \cite{Caffarelli_2007}. The physics of field theories with fractional Laplacian was also recently investigated from the point of view of entanglement entropy in \cite{Basa:2019ywr} (see also \cite{Nezhadhaghighi:2014pwa}), where it was found that the non-locality of this class of theories is responsible for the modification of the behaviour of entanglement entropy. 

Aside from entanglement entropy, long-range quantum systems have recently become interesting from the point of view of experiments involving cold atoms \cite{PhysRevLett.103.120502}. Entanglement of such systems has also been studied \cite{PhysRevLett.109.267203}, along with entanglement \cite{PhysRevA.89.062330} and the quench dynamics of entanglement \cite{PhysRevLett.113.210402}. Studies on entanglement entropy in Lifshitz type of theories have been done in \cite{MohammadiMozaffar:2018vmk, Mozaffar:2021nex}. One of the factors that make the quench in long-range systems interesting is the lack of the Lieb-Robinson velocity \cite{Hastings:2005pr, nachtergaele2006propagation}. The Lieb-Robinson velocity, which can be related to causality, governs the spread of a perturbation in a short-range system. Specifically, there is a causal region, outside which the perturbations drop off exponentially. Physically, this is equivalent to the presence of a velocity (the Lieb-Robinson velocity), which defines a cone-like causal region, in effect acting as the speed of light in these theories. Recently, there have also been a number of works studying the structure of light cones and the Lieb-Robinson velocity in long range systems (see, for example, \cite{foss2015nearly, gong2014persistence, guo2020signaling}).

The present paper is organised as follows. In section \ref{sec:nonlocaldesc}, the theoretical set-up of the model is described, along with the method used to calculate entanglement numerically. In section \ref{sec: static}, the results for the calculation of logarithmic negativity in the static case are given for both the critical and the non-critical harmonic chains and the effect of the long range interaction is discussed. Subsequently, in section \ref{sec: quench}, the phenomenon of quantum quench is described, along with the type of quench considered in this paper. The quasiparticle model of understanding post-quench evolution of the entanglement is also briefly introduced. Finally, the numerical results concerning the temporal evolution of logarithmic negativity are presented, with the lack of revivals in long range regimes emphasised. At the end, section \ref{sec: entanglement_entropy} describes similar results in entanglement entropy, with confirmation from the quasiparticle model alluded to earlier. The paper concludes with a summary of results and pointing out some future directions. Since there are indications of scrambling present in the results for logarithmic negativity and entanglement entropy, the calculation of the out-of-time-ordered correlators (OTOC) to probe this aspect is presented in the appendix.

\section{Non-local field theories with fractional Laplacian}
 \label{sec:nonlocaldesc}

\subsection{The Model}
\label{subsec: setup}

Some classes of non-local field theories considered in this paper have been studied from the point of view of entanglement entropy in \cite{Shiba:2013jja}. A slightly modified form of non-local scalar field theories with long-range interaction have also been investigated in \cite{Nezhadhaghighi:2014pwa} from the point of view of a quantum quench. In the present paper, logarithmic negativity will be calculated for non-local scalar field theories with fractional Laplacian, which is a non-local operator (see, for example, \cite{Frassino:2019yip}).

The class of theories under consideration are described by the Hamiltonian \cite{Nezhadhaghighi:2014pwa, Frassino:2019yip},
\begin{equation}
\label{Halpha}
H = \frac{1}{2}\int dx \left[ \pi^2 + \left( -\phi (-\Delta)^{\alpha/2} \phi \right) + m^{\alpha} \phi^2 \right].
\end{equation}
It may be observed that the above Hamiltonian generalises the Klein-Gordon scalar field theory by inclusion of the exponent, $\alpha$ of the Laplacian. The parameter $\alpha$ can take value greater than zero, with $\alpha = 2$ leading to the standard Klein-Gordon theory. The Hamiltonian can be discretised in the following form,
\begin{equation}
\label{Hrealspace}
H = a H_0 = \sum_{i=1}^{L} \frac{1}{2} \pi_i^2 + \sum_{i, j=1}^{L} \frac{1}{2} \phi_i V_{ij} \phi_{j}.
\end{equation}
Here, $a$ is the lattice spacing, which will henceforth be set to unity. For convenience of computations, the coupled system of oscillators obtained in eq. (\ref{Hrealspace}) can easily be decoupled by Fourier transformation to assume the form,
\begin{equation}
H = \sum_{k = 1}^{L}\omega_k(a^{\dagger}_k a_k + \frac{1}{2}),
\end{equation}
where the dispersion relation is given by,
\begin{equation}
\label{disprel}
\omega_k = \sqrt{m^{\alpha} + \left(4 \sin^2(\frac{\pi k }{L}) \right)^{\alpha/2}}.
\end{equation}

\subsection{Correlator method of calculating entanglement}

The subsequent sections of the paper are concerned with calculating entanglement in field theories in both static and quantum quench (time dependent) scenarios. The correlator method is a standard method used for calculating entanglement measures such as entanglement entropy and logarithmic negativity (see, for example \cite{serafini2017quantum}). Since this is discussed extensively in the literature, only a brief description is given here. For generality, the relevant expressions for the time dependent case will be given, which may readily be reduced to the corresponding quantities for the static scenario in the appropriate limit. To calculate entanglement measures, consider the time dependent correlators,
\begin{eqnarray}
Q_{ij}(t) &\equiv \langle{\psi} | \phi_i \phi_j | \psi \rangle = \frac{1}{2L} \sum_{i=1}^{L} Q_k \cos \left[  \frac{2\pi(i-j)k}{L} \right] ,\\
P_{ij}(t) &\equiv \langle{\psi } | \pi_i \pi_j | \psi  \rangle = \frac{1}{2L} \sum_{i=1}^{L} P_k \cos \left[  \frac{2\pi(i-j)k}{L} \right],\\
R_{ij}(t) &\equiv \langle{\psi} |\frac{1}{2} \{\phi_i ,\pi_j\} | \psi \rangle = \frac{1}{2L} \sum_{i=1}^{L} R_k \cos \left[  \frac{2\pi(i-j)k}{L} \right],
\end{eqnarray} 
and the ``correlation matrix" $\Gamma_{A \cup B}$ and the symplectic matrix $J$ which are written as,
 \[
\Gamma_{A \cup B} =
\begin{bmatrix}
Q_{ij}(t) & R_{ij}(t)  \\
R_{ji}(t) & P_{ij}(t) 
\end{bmatrix},
~~~~
J = \begin{bmatrix}
0 & I_{l\times l} \\
-I_{l \times l} & 0
\end{bmatrix}.
\]
The indices $(i,j)$ of the correlators runs over the regions included in the subsystem (whether adjacent or disjoint). For the quench protocol described above, the 2-point correlators are given by,
\begin{eqnarray}
\label{correlators}
Q_k &\equiv& \frac{1}{4 \omega_{1k} \omega_{2k}^2} \left(\omega_{1k}^2 + \omega_{2k}^2 - (\omega_{1k}^2 - \omega_{2k}^2)\cos(2\omega_{2} t) \right) ,\\
P_k &\equiv& \frac{1}{4\omega_{1k}} \left( \omega_{1k}^2 + \omega_{2k}^2 + (\omega_{1k}^2 - \omega_{2k}^2)\cos(2\omega_{2k} t) \right) ,\\
R_k &\equiv& \frac{\omega_{1k}^2 - \omega_{2k}^2}{4\omega_{1k} \omega_{2k}} \sin(2\omega_{2k} t).
\end{eqnarray}
Here, $\omega_{1k}$ and $\omega_{2k}$ are the pre and post-quench dispersion relations. In the quench scenario considered in this paper, the mass is abruptly changed. For generality, the expressions for the 2-point correlators are recorded above. The corresponding expressions for the static situation may be obtained by putting $\omega_{1 k} = \omega_{2 k}$ in the above expressions. 

It may be shown that the trace of the reduced density matrix $\textrm{Tr} \rho_A$ (where $A$ is the subsystem under consideration) can be related to eigenvalues ($\pm \lambda$) of the matrix $i J.\Gamma$ by the relation,
\begin{equation}
\textrm{Tr} \, \rho^n = \prod_{i\,=\,1}^{\ell}
\left[
\bigg( \lambda_i +\frac{1}{2} \bigg)^n - \bigg( \lambda_i-\frac{1}{2} \bigg)^n\,
\right]^{-1}.
\end{equation}
The entanglement entropy is then given by,
\begin{equation}
S_A (t) = \sum_{i = 1}^{\ell} (\lambda_{i} + \frac{1}{2}) \log(\lambda_{i} + \frac{1}{2}) - (\lambda_{i} - \frac{1}{2}) \log(\lambda_{i} - \frac{1}{2}),
\end{equation}
where the size of the subsystem concerned is given by $\ell$. To calculate the logarithmic negativity, consider a region $(A \cup B)$ of the system and partial transpose with respect to $B$. In the correlation matrix $\Gamma$, the partial transpose merely inverts the signs of the matrix elements corresponding to momenta included in $B$. So the partial transpose of the matrix $\Gamma$ is given by,
\begin{equation}
\Gamma_{A \cup B}^{T_2} = \begin{pmatrix}
\mathbb{I}_{\ell} & \boldsymbol{0}_{\ell}  \\
\boldsymbol{0}_{\ell}  & \mathbb{R}_{B}
\end{pmatrix}
\cdot 
\Gamma_{A \cup B}
\cdot 
\begin{pmatrix}
\mathbb{I}_{\ell} & \boldsymbol{0}_{\ell}  \\
\boldsymbol{0}_{\ell}  & \mathbb{R}_{B}
\end{pmatrix}.
\end{equation}
Finally the trace norm of the partially transposed density matrix $\rho_{A \cup B}^{T_2}$ which is needed for calculation of negativity is given as,
\begin{equation}
|| \rho^{T_2} || 
= 
\prod_{i\,=\,1}^{\ell}
\Bigg[\,
\bigg| \chi_i+\frac{1}{2} \bigg| - \bigg| \chi_i -\frac{1}{2} \bigg| \,
\Bigg]^{-1}
= 
\prod_{i\,=\,1}^{\ell}
\textrm{max}
\bigg[ 1 ,  \frac{1}{2\chi_i} \bigg]\,,
\end{equation}
where $\chi \equiv \chi(t)$ is the set of eigenvalues of the matrix $i J.\Gamma_{A}^{T_2}$. Finally, the logarithmic negativity of the subsystems $A$ and $B$ may be calculated from eqn.
\begin{equation}
\label{negvmaster}
E_{LN}(A, B) \equiv \lvert \lvert \rho_{A \cup B}^{T_2}  \rvert \rvert,
\end{equation}
by substituting the expression of $|| \rho_A^{T_2} || $.

The mutual information similarly may be obtained simply by calculating,
\begin{equation}
I(A, B) = S(A) + S(B) - S(A \cup B).
\end{equation}

\section{Static scenario: Results for the logarithmic negativity}
\label{sec: static}

The simplest physical scenario to investigate the effect of the long range interactions (controlled by the parameter $\alpha$) on logarithmic negativity is the static situation. This would serve as a warm-up to understand better the behaviour of logarithmic negativity in more complex scenarios (such as the quantum quench considered in subsequent sections). Since the regime $\alpha>2$ corresponds to short range models (with $\alpha = 2$ corresponding to the Klein-Gordon case), the range of $\alpha$ is confined to $0 < \alpha <2$ in the remainder of this section. The analysis is done for the critical ($m \approx 0$)\footnote{In practice, a small value of $m=10^{-5}$ is chosen to ensure numerical stability.} and the non-critical cases, similar to \cite{Marcovitch:2008sxc}, which investigated similar issues for the Klein-Gordon field. As with the rest of the present work, the results are presented for a finite size system, i.e, for a system in which the ratio $\ell/L$ is non-zero and finite. 

\begin{figure}[htp]{
		\begin{tabular}{cc}
			\includegraphics[width=0.5\textwidth]{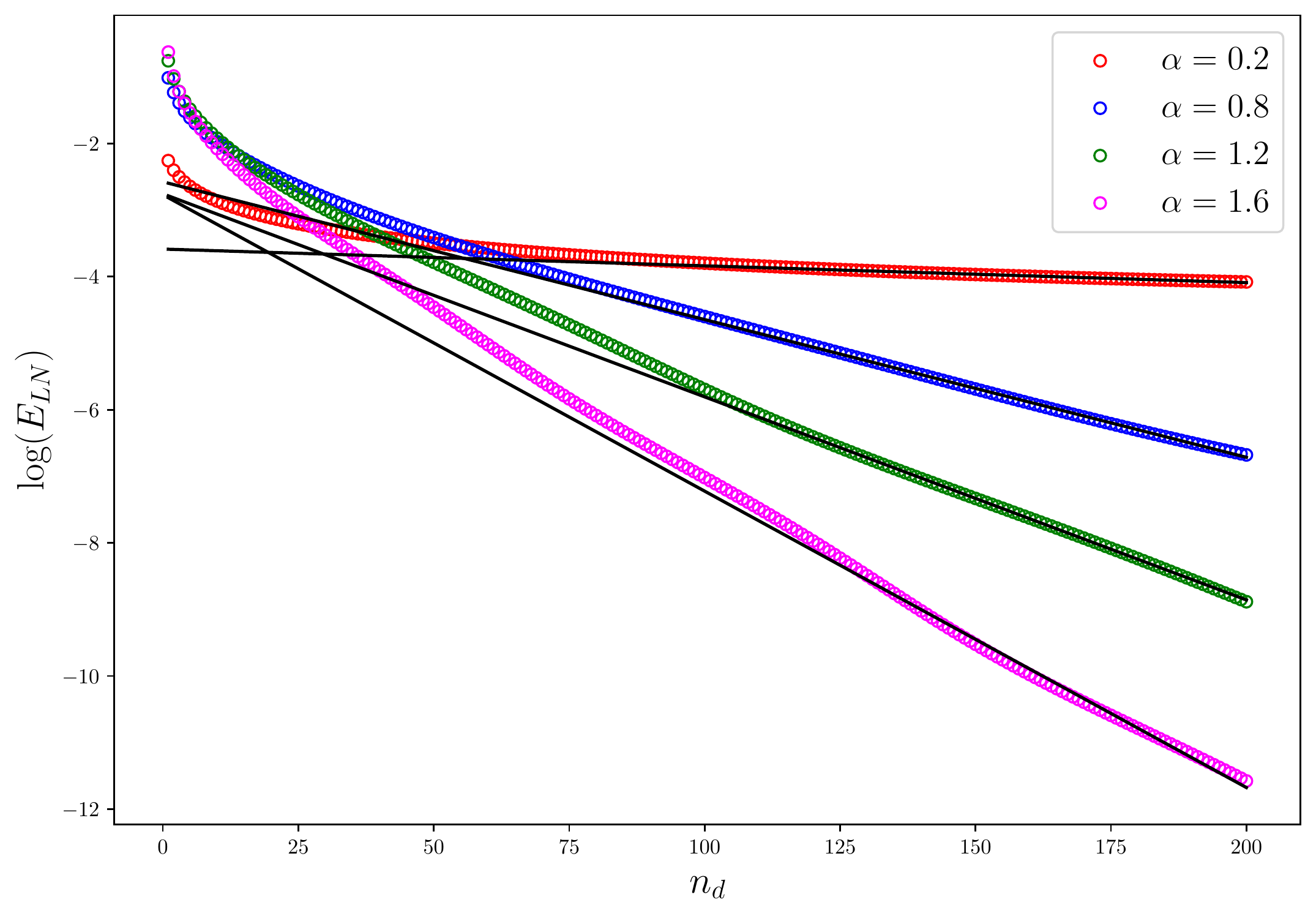}&
			\includegraphics[width=0.5\textwidth]{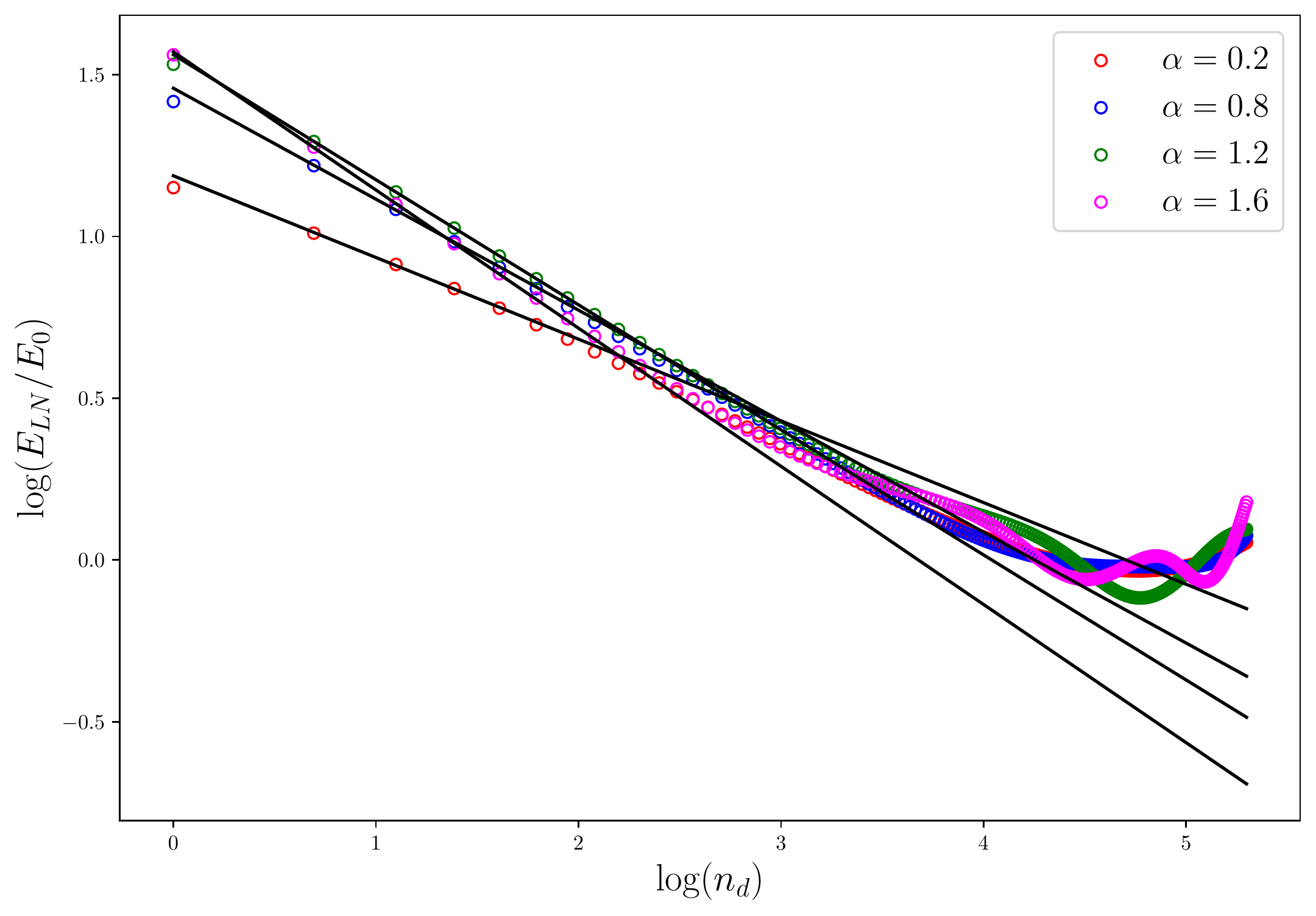}\\
			(a) & (b)
		\end{tabular}
		\caption{Critical: (a) $\log E_{LN}$ vs $n_d$ for different values of the parameter $\alpha$. (b) $\log E_{LN}$ as a function of $\log n_d$. In all cases, the black lines denote a linear fit. The total number of sites is $L = 1250$ and the subsystem size is $n_A = n_B = 50$ with $m = 10^{-5}$.} 

\label{static: critical}}
\end{figure}

As mentioned earlier, the logarithmic negativity is calculated by means of the correlator method, with the pre and post-quench masses set equal in the general (time-dependent) expressions for the 2-point correlators in eq. (\ref{correlators}). In Fig. \ref{static: critical}, for the critical case the results are presented, focusing on the fall-off the entanglement as the separation between the subsystems is increased.

In Fig. \ref{static: critical}, it is of interest to check regions of exponential and/or power law fall-off in the negativity. Consequently,  for Fig. \ref{static: critical}(a), a linear fit is done at large separation ($n_d > 60$ in this particular case) and at small separation ($n_d < 2.5$) for Fig. \ref{static: critical}(b). From Fig. \ref{static: critical}, it is clear that, the fall-off is power-law at short distances and changes to an exponential fall-off at larger distances. It is also clear especially from Fig. \ref{static: critical}(a) that the linear approximation (corresponding to exponential decay) appears to break down for increasing values of $n_d$ as $\alpha$ increases, i.e., the fall-off of $E_{LN}$ assumes an exponential form at increasingly smaller separations of the subsystems when the the long range interaction increases. It may be noted that the slope of the linear fit in Fig. \ref{static: critical}(b) depends on the parameter $\alpha$.\footnote{For the critical Klen-Gordon field and for critical spin chain, the power-law exponent has been determined to be $1/3$, which is the same as the prefactor to the entanglement entropy \cite{Marcovitch:2008sxc}, \cite{Wichterich:2008vfx}.}

\begin{figure}[htp]{
		\begin{tabular}{cc}
			\includegraphics[width=0.5\textwidth]{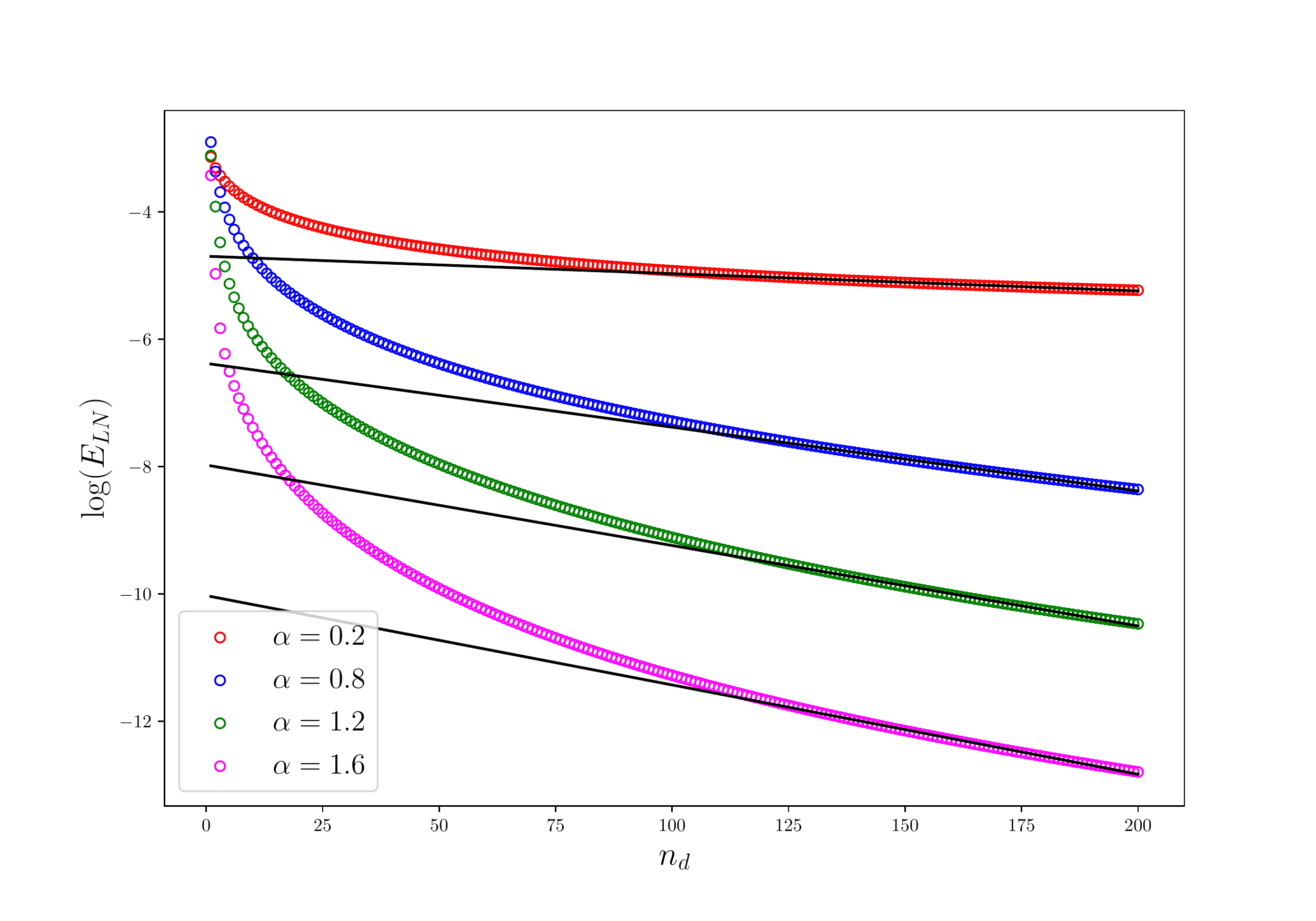}&
			\includegraphics[width=0.5\textwidth]{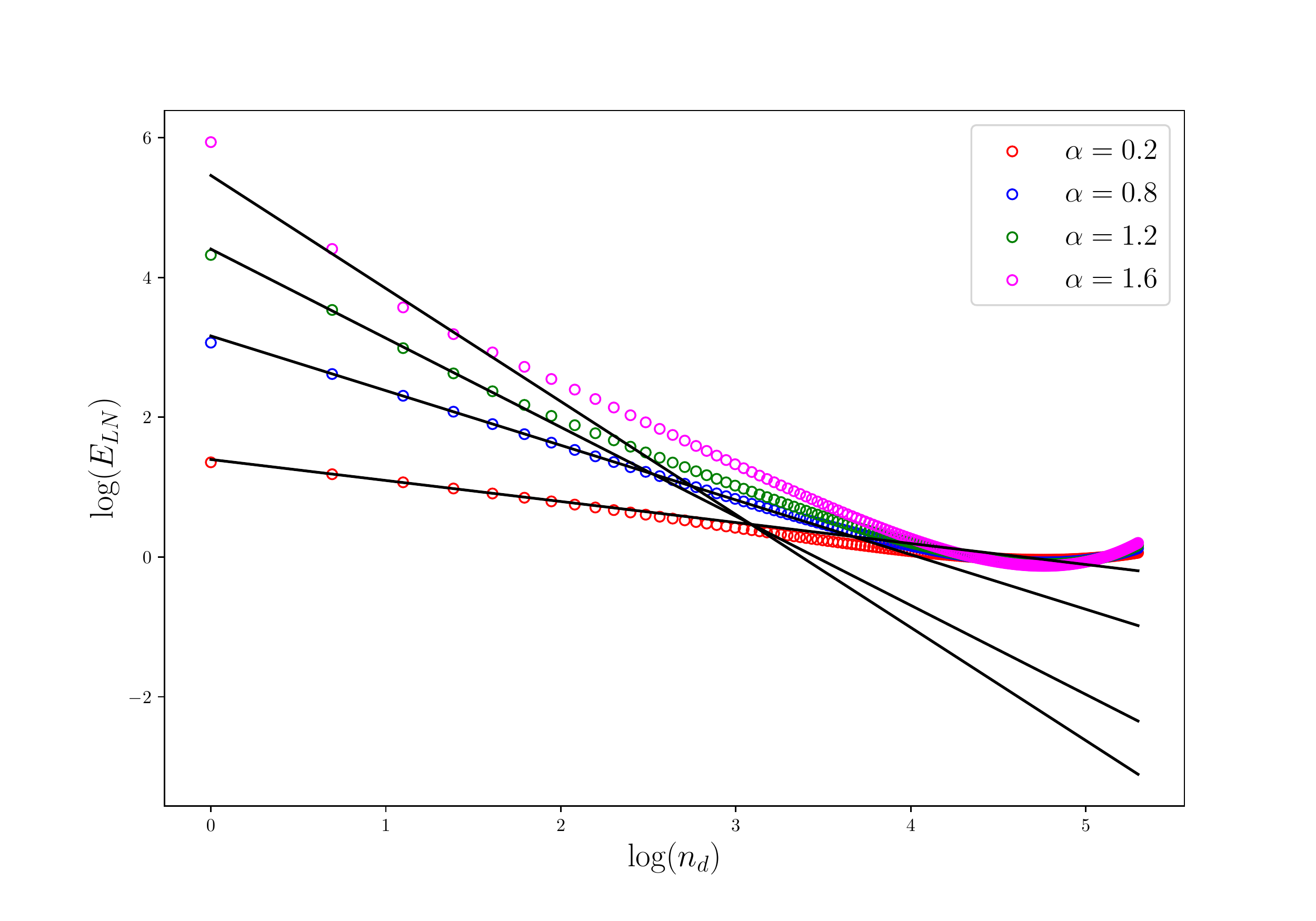}\\
			(a) & (b)
		\end{tabular}
		\caption{Non-Critical: (a) $\log E_{LN}$ vs $n_d$ for different values of the parameter $\alpha$. (b) $\log E_{LN}$ as a function of $\log n_d$. In all cases, the black lines denote linear fit. The total number of sites is $L = 1250$ and the subsystem size is $n_A = n_B = 50$ with $m = 1.0$.} 
	
	\label{static: noncritical}}
\end{figure}

For the non-critical case, it is also desirable to  study the variation of $E_{LN}$ as a function of the separation $n_d$ between the subsystems, the results of which are shown in Fig. \ref{static: noncritical}. Comparing Figs. \ref{static: critical} and \ref{static: noncritical}, it is clear that the deviation from exponential fall-off behaviour is more pronounced for the non-critical case, i.e, the power law fall-off persists for larger distances.

\begin{figure}[htp]{
			\begin{tabular}{cc}
		\includegraphics[width=0.5\textwidth]{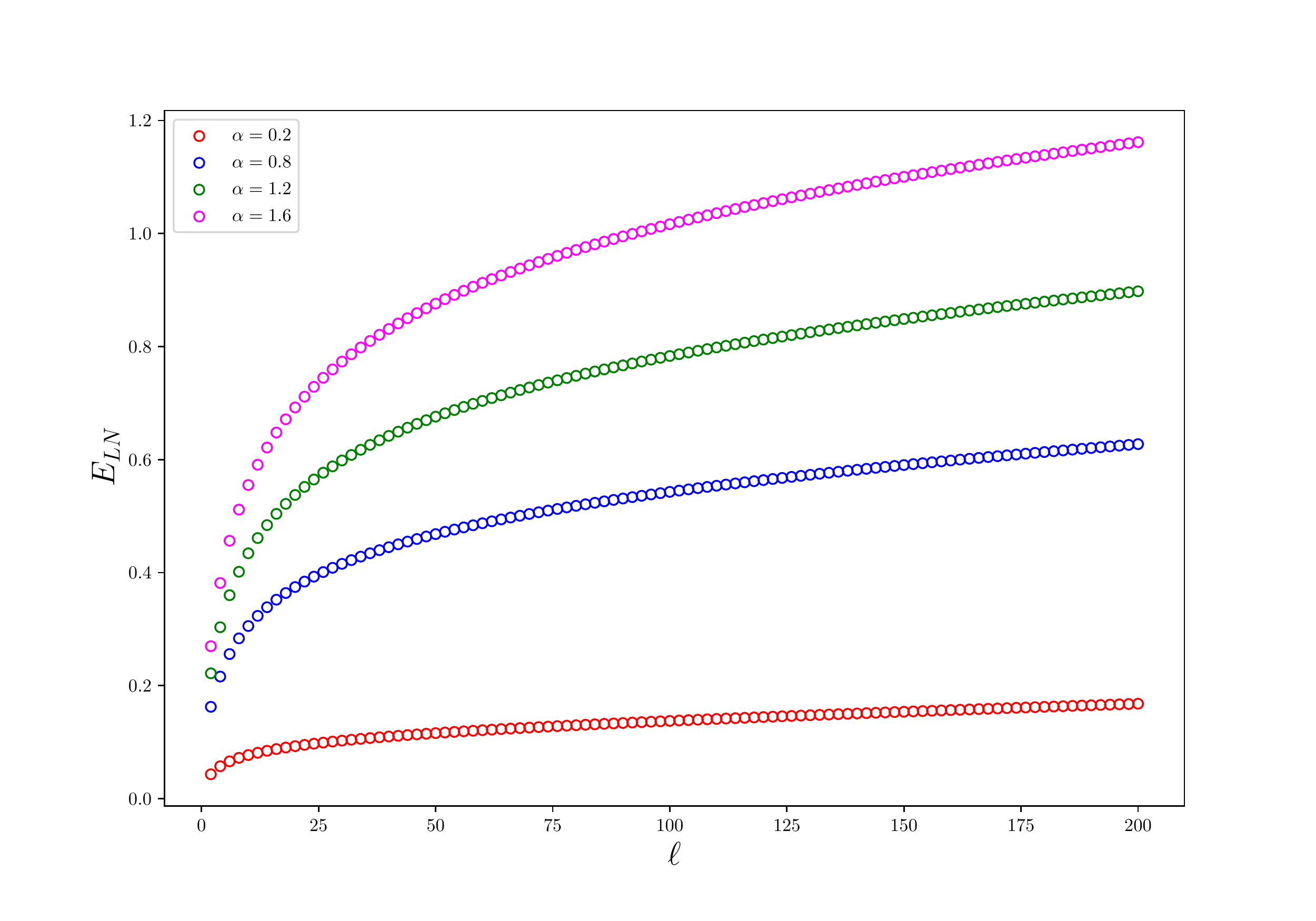}&
		\includegraphics[width=0.5\textwidth]{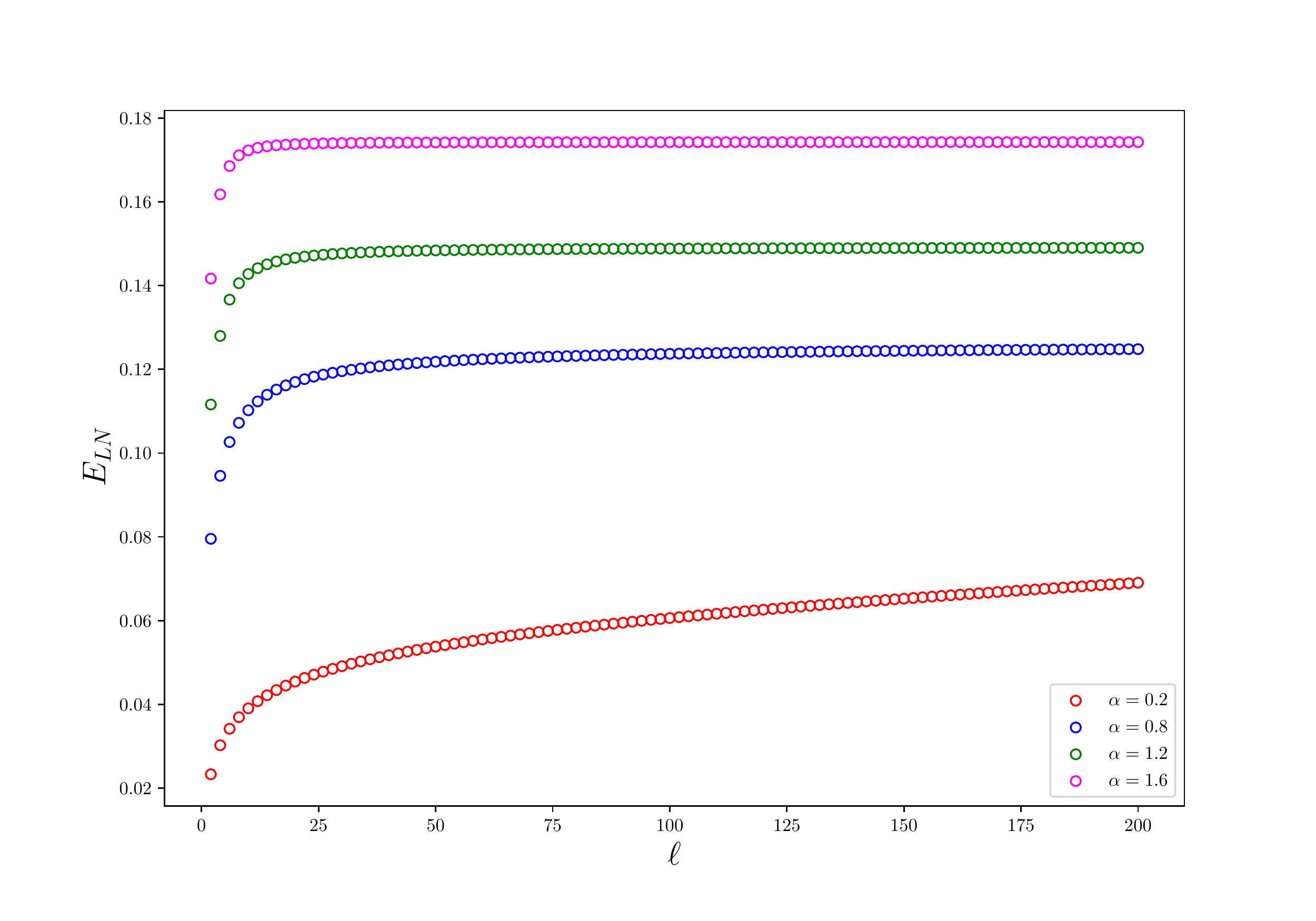}\\
		(a) & (b)
	\end{tabular}
\caption{(a) Critical: $E_{LN}$ vs $\ell$ for $m = 10^{-5}$. (b) Non-critical: $E_{LN}$ vs $\ell$ for $m = 1.0$. In both cases, total number of sites $N = 1250$ and $n_A = n_B = 50$.}

\label{static: lgnvsl}}
\end{figure}

It is also interesting to plot the logarithmic negativity with respect to subsystem size, for the critical and non-critical cases (Fig. \ref{static: lgnvsl}). From the Fig \ref{static: lgnvsl}, it is clear that the critical case can be distinguished from the non-critical by the observation that $E_{LN}$ saturates rapidly for the non-critical case compared to the critical case.
	
\section{Quantum quench scenario}
\label{sec: quench}

With the results of the previous section which detail the general properties of logarithmic negativity in a static scenario, it is suitable to analyse the temporal evolution of logarithmic negativity in a time dependent scenario. A quantum quench is defined as a physical situation in which a parameter in the Hamiltonian undergoes a change with respect to time. In the specific protocol for the quench considered in this paper, the mass parameter in the Hamiltonian of the system (eq. (\ref{Halpha})) is considered to change suddenly. To gain physical understanding of the behaviour of the system post-quench, it is useful to recall a heuristic model of a quantum quench that exists in the literature. 

\subsection{The quasiparticle picture for quantum quench}
\label{subsec: quasipic}
It is well known that the post-quench behaviour of a measure of entanglement may be understood in qualitative (and in some cases quantitative) terms by a simple semiclassical model of the quench \cite{Calabrese:2005in}. This model has been applied successfully to study quenches in various systems like conformal field theory and many-body systems \cite{Fagotti:2010yr}. Although it is difficult to apply this model to complicated entanglement measures like negativity, it is sufficient to describe measures like entanglement entropy well, which would serve to provide some indication of the physics in other scenarios also. 
 
Considering the energy difference between the initial and the post-quench state, the quench results in generation of quasiparticle excitations in pairs, which then propagate through the system and generate entanglement. However, to contribute to entanglement, the quasiparticle pair should be generated at the same point in the system. Specifically, entanglement between two regions (say, $A$ and $B$) is generated by one member of the quasiparticle pair arriving at $A$, while the other member arrives at $B$ simulteneously.

It has been proposed in  \cite{Calabrese:2005in} that the entanglement is proportional to the number of quasiparticles, resulting in the relation (for a subsystem of size $\ell)$,
\begin{equation}
S_{\ell}(t) \propto 2t \int_{2|v|t < \ell} dk v(k)s(k) + \ell \int_{2|v|t>\ell} dk s(k) .
\end{equation}
Above, $v(k)$ refers to the velocity of propagation of the quasiparticles $s(k)$ indicates the production rate of quasiparticles and the entropy density.  According to \cite{Calabrese:2005in}, $s(k)$ refers to the contribution of quasiparticles with momenta $\pm k$ to the thermodyanamic entropy of the Generalized Gibbs Ensemble (GGE) of the steady state. It can be observed from this equation that the existence of a maximum velocity of excitations ($v_M$) ensures a light-cone like spreading of correlations. It should be noted that the above equation is valid for the continuum limit. 

For systems with finite size, the above equation is modified to obtain \cite{Modak:2020faf},
\begin{multline}
S_{\ell}(t) = \int_{ \{ \frac{2v(k)t}{L} \} < \frac{\ell}{L} } \frac{dk}{2\pi} s(k) L \{\frac{2v(k)t}{L} \} + \ell \int_{ \frac{\ell}{L} \leq \{\frac{2v(k)t}{L} \} < 1 - \frac{\ell}{L}} \frac{dk}{2\pi}s(k) +  \\
\int_{1 - \frac{\ell}{L} \leq \{ \frac{2v(k)t}{L} \}} 	\frac{dk}{2\pi}	s(k)L \left( 1 - \{ \frac{2v(k)t}{L} \} \right) .		
\end{multline}
In the above equation, $\{x \}$ indicates fractional part of $x$. A feature of the above equation is that ``revivals'' in entanglement are accounted for, which also occur in numerical results for finite systems (for example, in \cite{Coser:2014gsa}). 

At this point, it is worthwhile writing down the exact expressions for the specific class of models which occur in this paper. Firstly, the velocity of propagation of the excitations is given by,
\begin{equation}
v(k) = \frac{d\omega_k}{dk} = \frac{2^{-2+a} a \cot(\pi k/L) (\sin(\pi k/L))^a}{\sqrt{m_2^a + (4 \sin(\pi k/L)^2)^{a/2}}}  ,
\end{equation} 
where $m_2$ represents the post-quench mass. The GGE density matrix can be constructed from the creation/annihilation operators as,
\begin{equation}
\rho_{\textrm{GGE}} = \frac{e^{-\sum_k \lambda_k \hat{n}_k}}{Z},
\end{equation}
where $Z = \textrm{Tr} e^{-\sum_k \lambda_k \hat{n}_k}$. Note that $\lambda_k$ denotes Lagrange multiplier, and they are evaluated from the constraint that $\langle \hat{n}_k \rangle = \langle \psi_0 | n_k | \psi_0 \rangle$ in the initial state must equal the value according to GGE,
\begin{equation}
n_k = \langle \hat{n}_k \rangle = \frac{1}{e^{\lambda_k}-1}.
\end{equation}
The thermodynamic entropy according to GGE is,
\begin{equation}
S_{\textrm{GGE}} = \sum_k s(k) = -\textrm{Tr}\left( \rho_{\textrm{GGE}} \log \rho_{\textrm{GGE}}\right) = \sum_k \left[ (n_k+1)\log(n_k+1) - n_k \log n_k \right]
\end{equation}
In case of, for example, the massive to massless quench, the occupation number $n_k$ can be easily calculated to be,
\begin{equation}
n_k = \frac{1}{4} \left( \frac{\omega_{2 k}}{\omega_1k} + \frac{\omega_{1 k}}{\omega_{2 k}} \right) - \frac{1}{2} . 
\end{equation}
Here, $\omega_{1 k}$ and $\omega_{2 k}$ are pre and post-quench dispersion relations respectively.

The 2-point correlators for these models have also been studied in \cite{PhysRevB.91.045131}, for both massless and massive post-quench Hamiltonians. It was established that for the critical quench, the ballistic velocity of propagation of the quasiparticles showed no maximum, corresponding to absence of Lieb-Robinson bound for all values of $\alpha$. For massive post-quench Hamiltonian, the velocity of propagation showed a maximum for $\alpha > 1.0$ and was unbounded for $\alpha<1$, with the former case corresponding to a generalised Lieb-Robinson bound.

\begin{figure}[htp]{
			\begin{tabular}{cc}
		\includegraphics[width=0.5\textwidth]{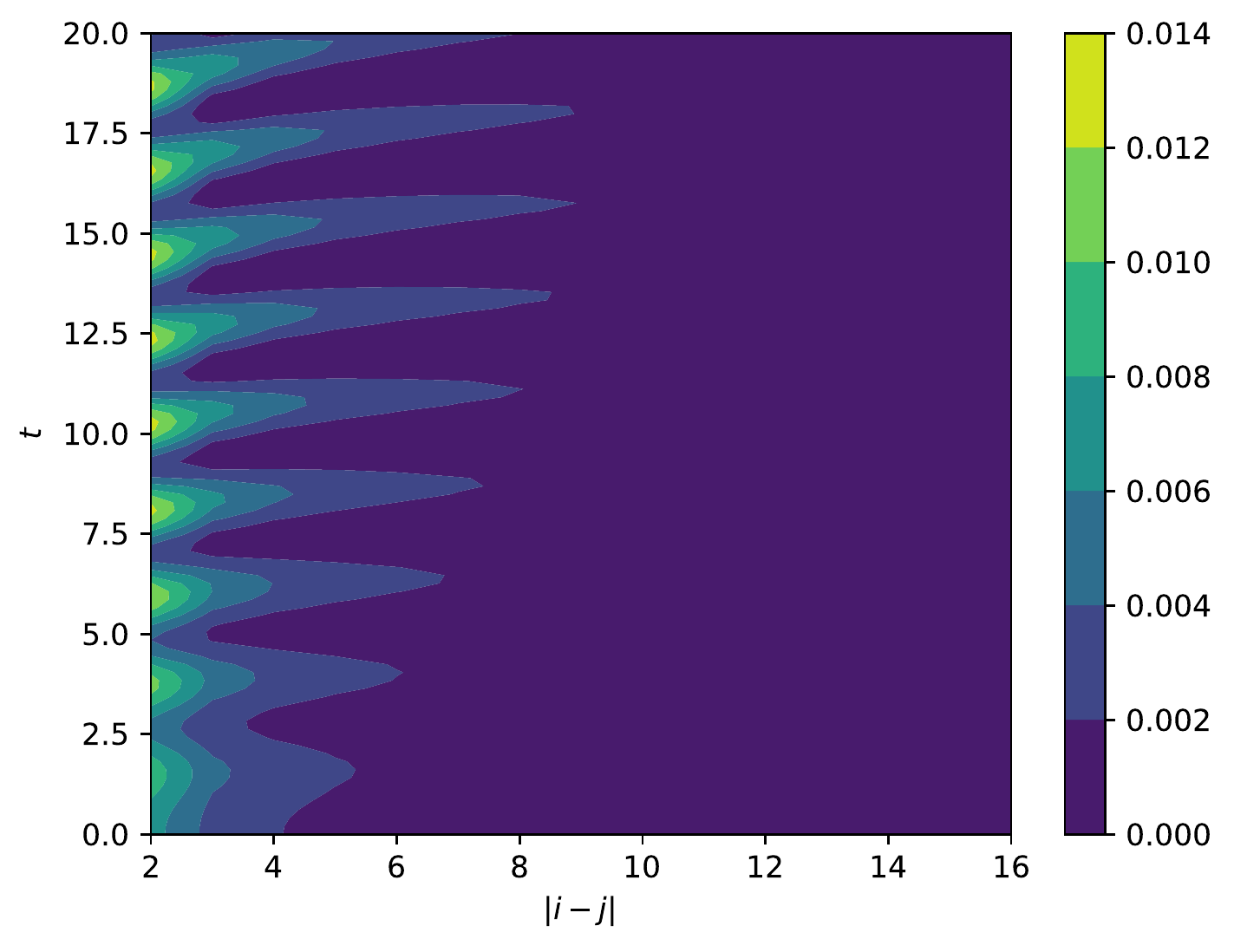}&
		\includegraphics[width=0.5\textwidth]{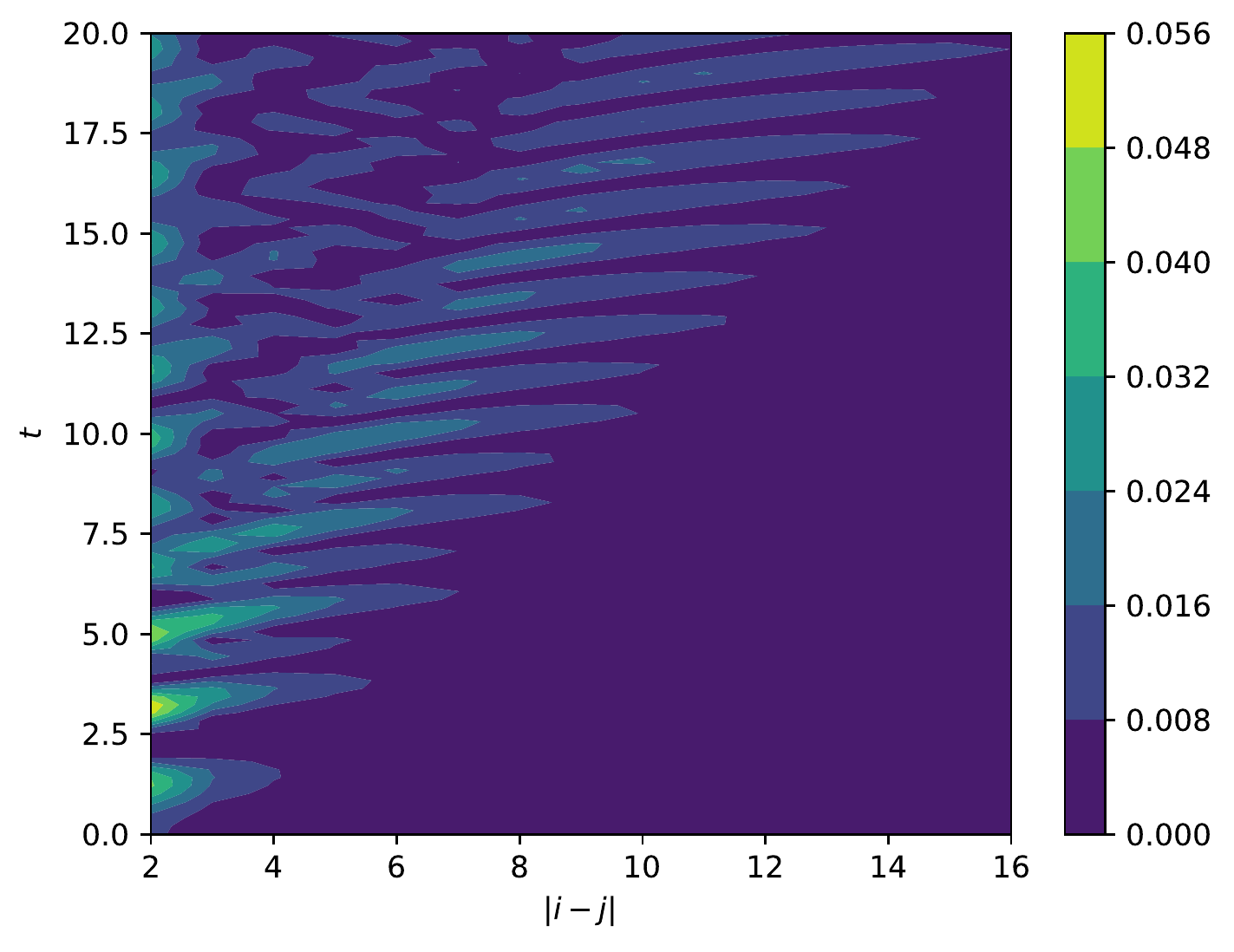}\\
		(a) & (b) \\
		\includegraphics[width=0.5\textwidth]{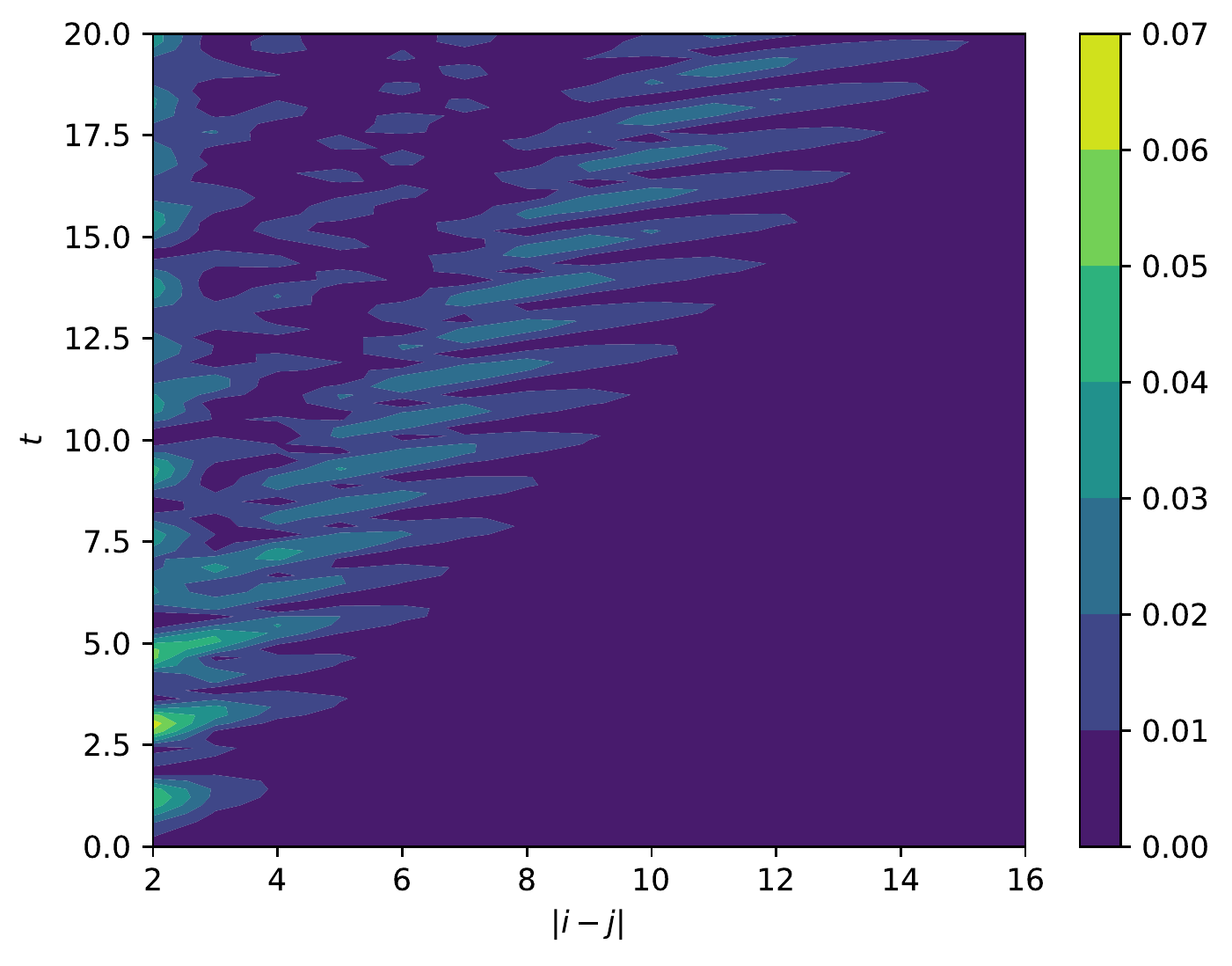}&
		\includegraphics[width=0.5\textwidth]{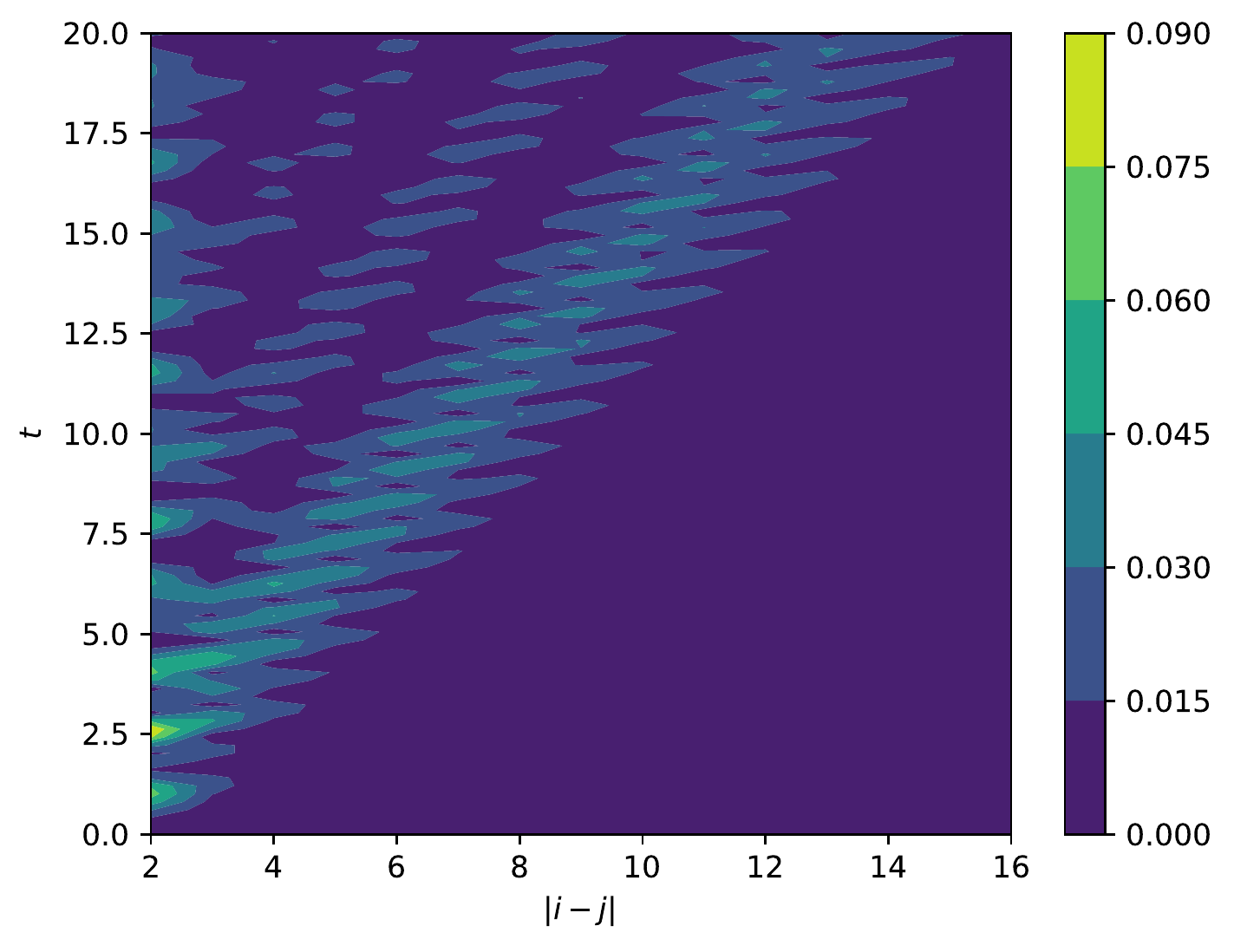}\\
		(c) & (d)
	\end{tabular}
\caption{Density plot for $|Q_{ij}|$ against $t$ and $|i-j|$ reveal the change in shape of the light cone and the range of interactions in the model is varied. Clockwise from the top, the values for $a$ considered are $a = 0.2, 0.9, 1.1, 1.6$. In each case, the total number of sites is $L = 1250$ and the quench protocol is the sudden change of $m=4$ to $m=2$.}
\label{quench: light_cones}}
\end{figure}

Fig. \ref{quench: light_cones} depicts the spread of information in a finite-sized system and it may be observed that the shape of the ``light cone'' undergoes a marked change as $a$ is increased. From the preceding description of the quasiparticle model for quenches, it is natural to expect that this would influence the temporal behaviour of the entanglment post-quench.

\subsection{Quantum quench scenario: Numerical results for logarithmic negativity}
In this subsection, numerical results for the logarithmic negativity are presented, considering a quench from $m = 1$ to $m = 0$. As mentioned earlier, the correlator method is used, with the relevant expressions for the individual correlators given by eqn. (\ref{correlators}).

\begin{figure}[htp]{
		\begin{tabular}{cccc}
			\includegraphics[width=0.5\textwidth]{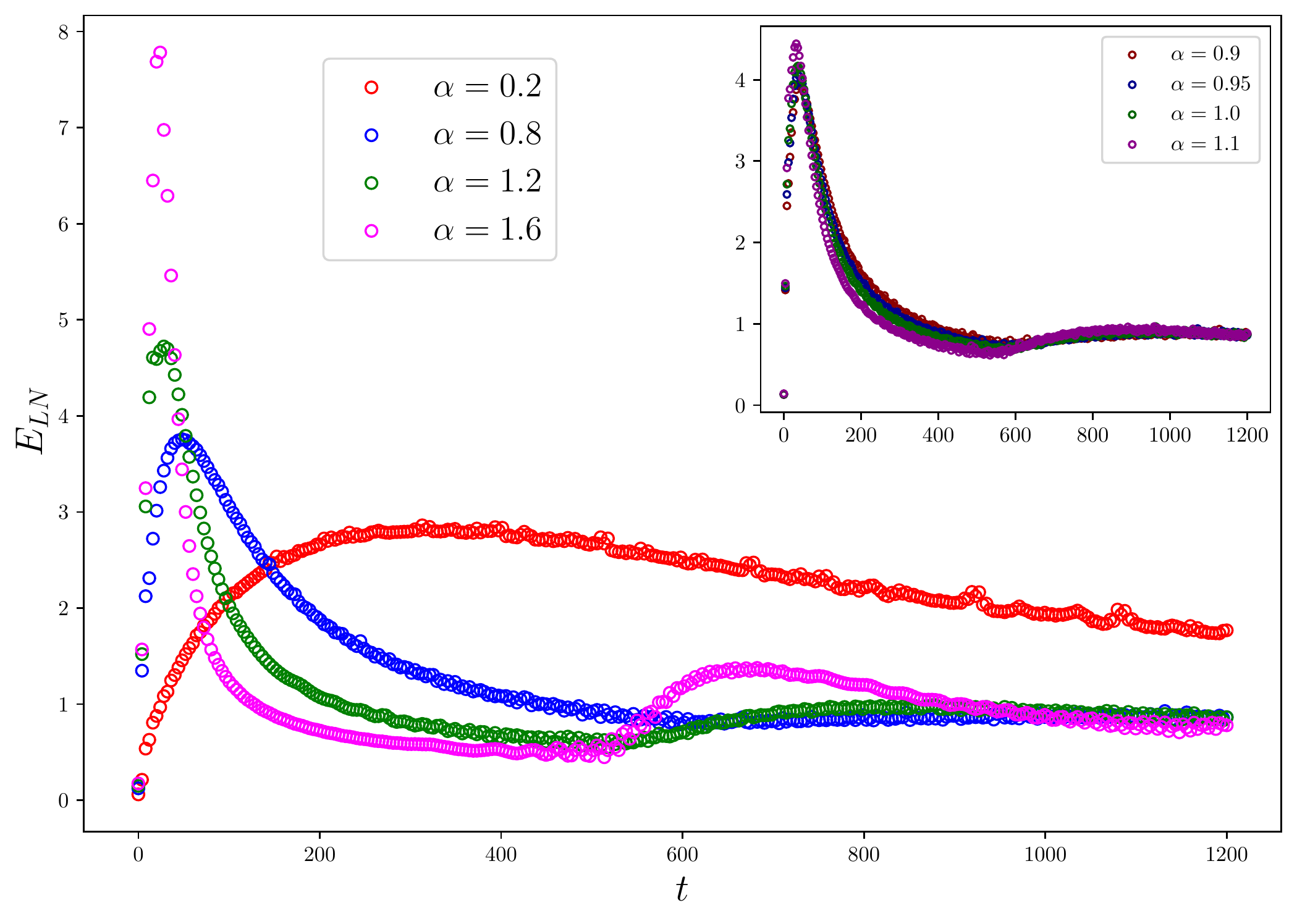}&
			\includegraphics[width=0.5\textwidth]{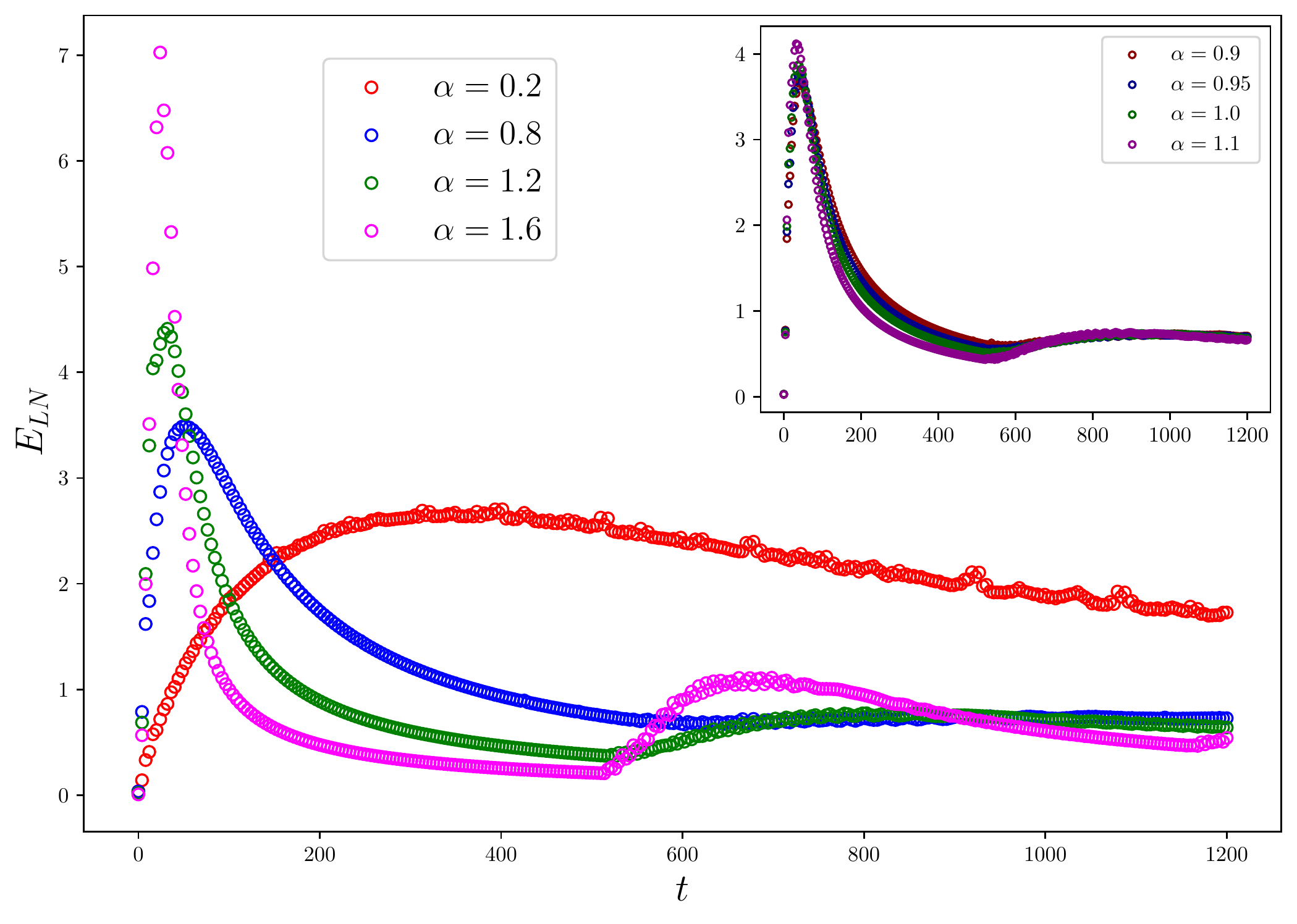}\\
			(a) & (b) \\
			\includegraphics[width=0.5\textwidth]{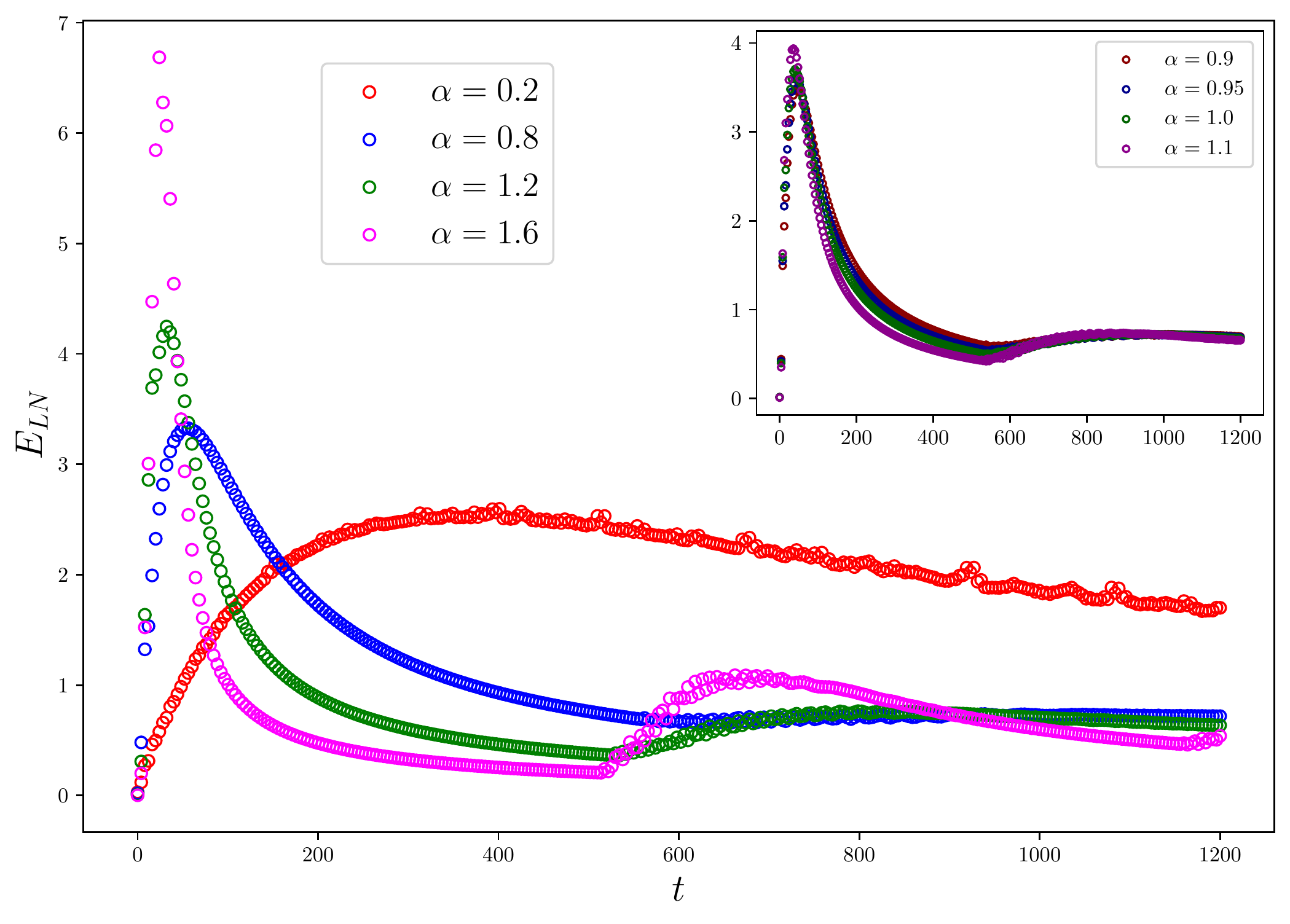} &
			\includegraphics[width=0.5\textwidth]{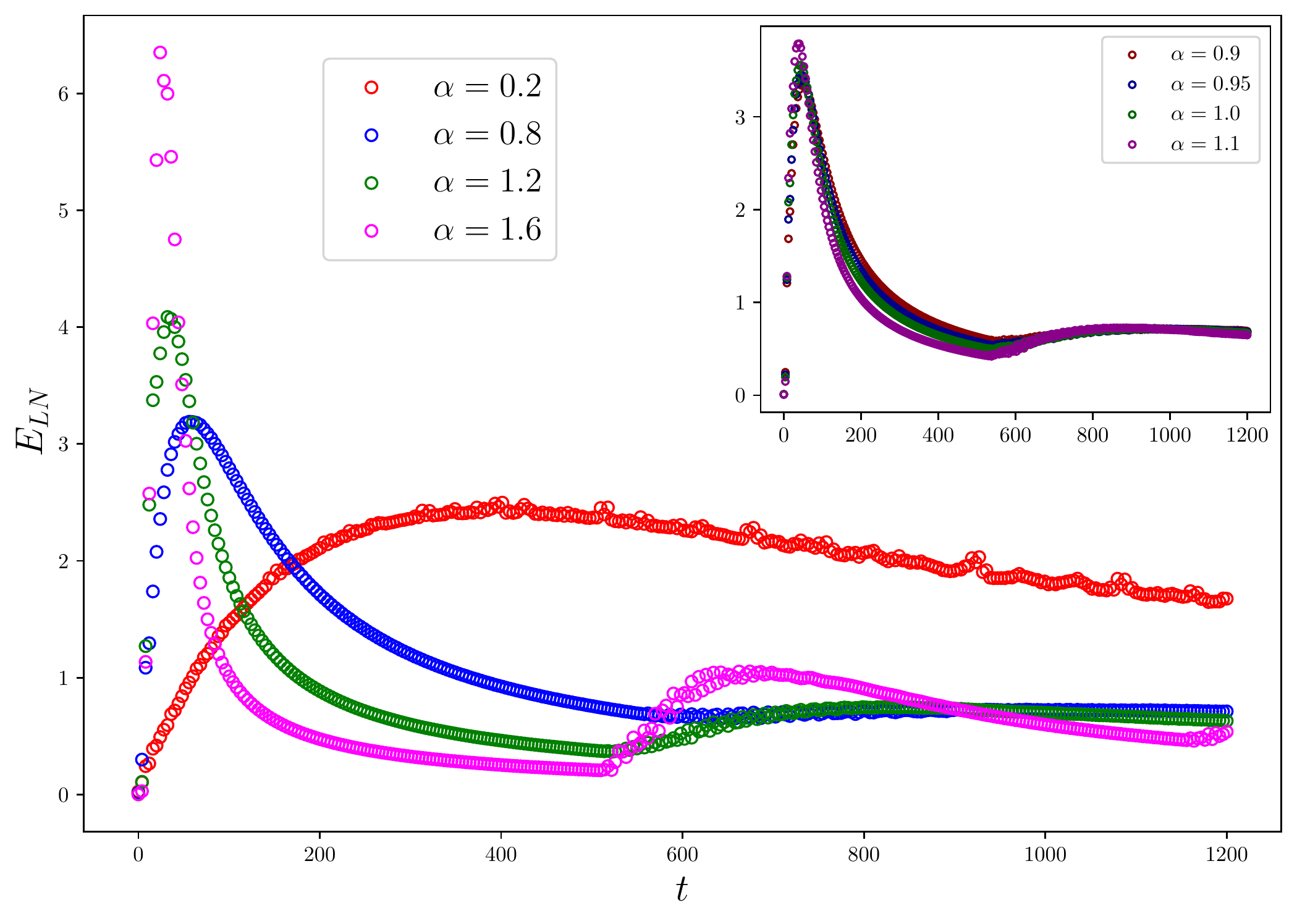} \\
			(c) & (d)
		\end{tabular}
		\caption{$E_{LN}$ vs $t$ for different values of the $\alpha$ and the separation between subsystems $n_d$. For (a), (b), (c) and (d), $n_d$ takes values $0$, $2$, $4$ and $6$ respectively. The inset(s) examine the limit $\alpha \rightarrow 1$ in more detail. In each case, $L = 1250$ and $n_A = n_B = 50$, with the mass parameter quenched from $m=1$ to $m=0$.}
	
	\label{critical_quench: lgnvst}}
\end{figure}
It is instructive to note how the qualitative behaviour of logarithmic negativity changes from $0<\alpha<1$ to $\alpha > 1$ in Fig. (\ref{critical_quench: lgnvst}). To emphasise the robustness of the results, several different cases, corresponding to different separations are examined and confirmed to reach the same conclusion. As can be observed, (partial) revivals are observed in each case as the limit $\alpha \rightarrow 1$ is crossed. Recalling the content of the previous section, this corresponds to the limit of the ballistic velocity of propagation of quasiparticles acquiring a bound. 

The inset of each figure examines more closely the limit $\alpha = 1.0$, at which it appears that qualitative changes to appear. Of course, it is reasonable to ask whether as $\alpha \rightarrow 1$ is crossed, the revivals are pushed out to $t = \infty$. It is difficult to confirm this entirely conclusively in a numerical study, but this issue is partly addressed in the subsequent section.

\begin{figure}[htp]{
\includegraphics[scale=0.7]{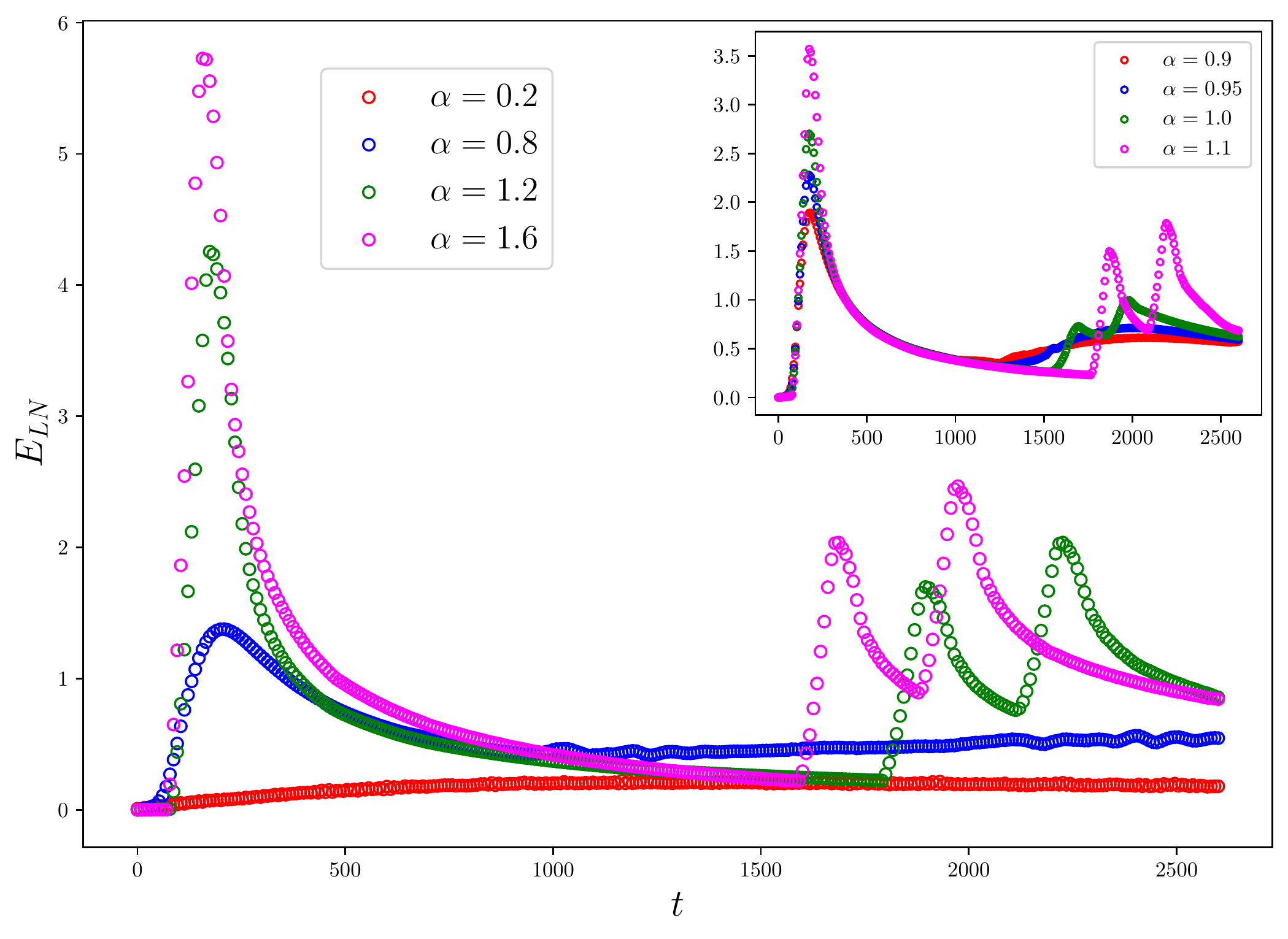}
\label{massive_quench: lgnvst}
\caption{$E_{LN}$ vs $t$ for different values of the $\alpha$ with the separation between subsystems $n_d=10$. Here $L = 1250$ and $n_A = n_B = 50$, with the mass parameter quenched from $m=4$ to $m=2$.}
\label{massive_quench: lgnvstmassive}}
\end{figure}
It is also relevant to check whether the post-quench Hamiltonian being massless affects the results. From Fig. \ref{massive_quench: lgnvstmassive} where the mass parameter is quenched from $m=4$ to $m=2$, this would appear to not be the case since revivals occur in this case also. It is clear that the revivals appear for $\alpha>1$, with the logarithmic negativity behaving in a qualitatively different manner for lower values of $\alpha$. From the results in this section, it is clear that revivals are present in logarithmic negativity in quantum quenches from $\alpha \approx 1$ onwards, thus establishing this as a general feature.

\section{Investigating revivals in entanglement entropy}
\label{sec: entanglement_entropy}

\begin{figure}[htp]{
\includegraphics[scale=1.0]{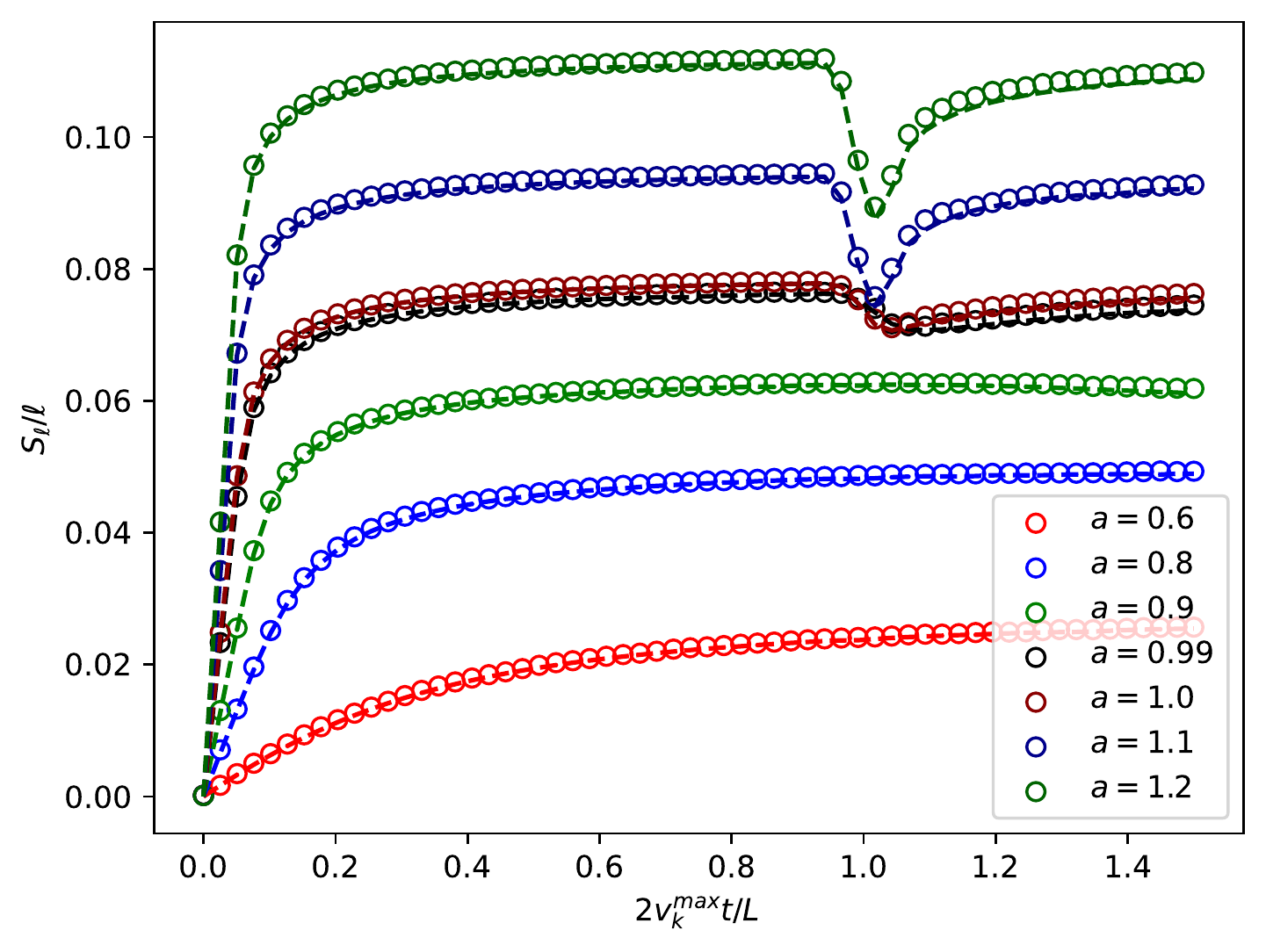}
\caption{$S_{\ell}/\ell$ vs $t$ for different values of the $\alpha$. Here $L = 1250$ and $n_A=50$, with the mass parameter quenched from $m=4.0$ to $m=2.0$. Here, the dotted lines correspond to the predictions from the quasiparticle model.}
\label{seevst}
}
\end{figure}

Based on the results in the previous section, it seems that gradual appearance of revivals as $\alpha$ is increased is a general feature in logarithmic negativity. As a check, it is desirable to confirm the general nature of revivals in another measure of entanglement, namely, entanglement entropy. Although entanglement entropy is a much simpler measure of entanglement, it would be interesting if it showed any notable/similar charactertistics in long range theories. This has the advantage that predictions can be checked against the quasiparticle model (described in section \ref{subsec: quasipic}).\footnote{To the author's knowledge, there is no proposal in the literature for calculating a complicated measure such as logarithmic negativity via the quasiparticle model.} The results are given in Fig. \ref{seevst}, which shows that a dip in entanglement occurs at $t = 2 v_k^{max}/L$ for $\alpha>1$. This is expected from the quasiparticle model and is also observed in \cite{Modak:2020faf}, which is the main motivations for this section. As in the logarithmic negativity, no revivals are observed for approximately $\alpha<1$. The results show good agreement with the predictions from the quasiparticle picture, denoted by the dashed lines in Fig. \ref{seevst}

\begin{figure}[htp]{
\includegraphics[scale=1.0]{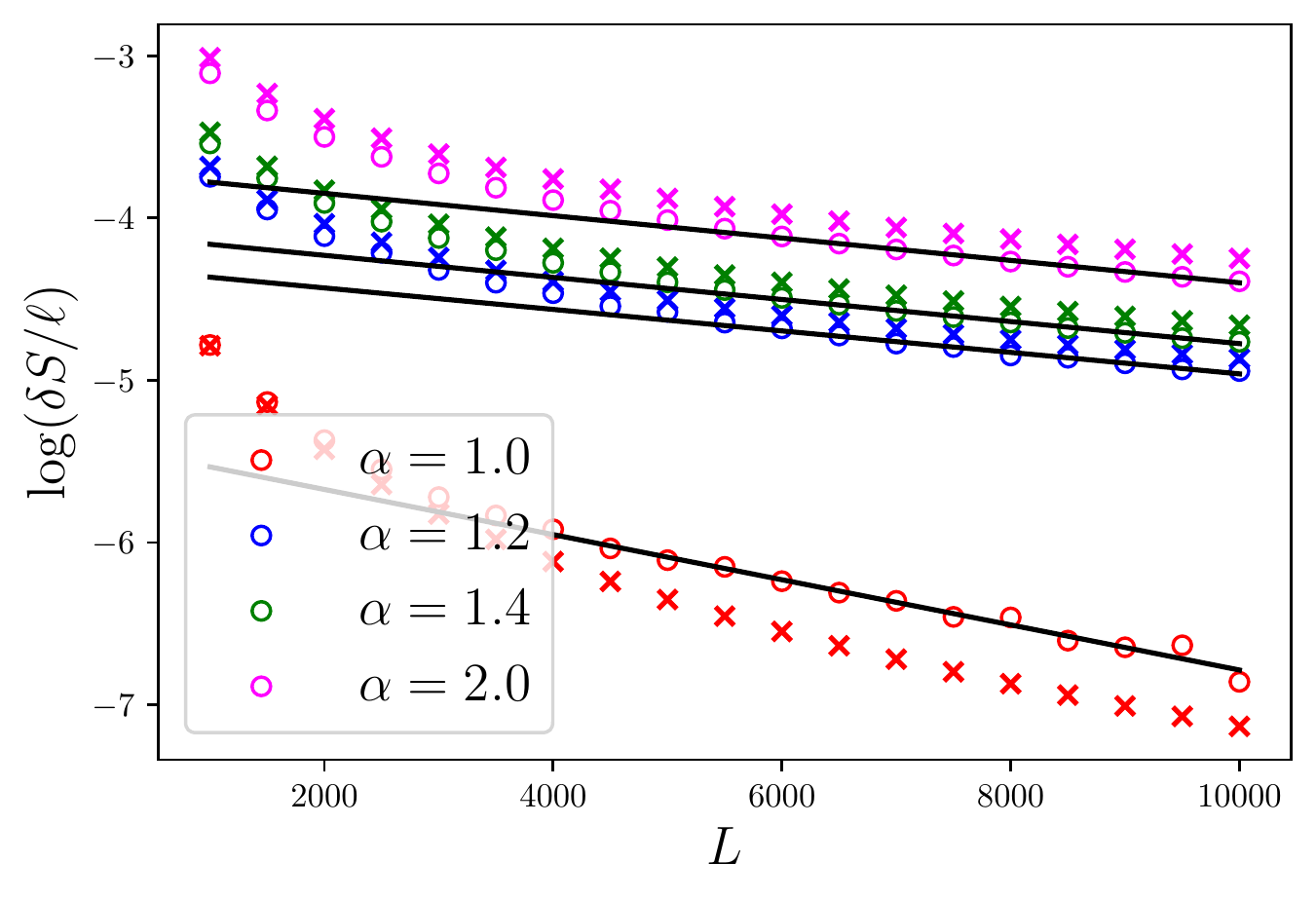}
\caption{The first dip in the entanglement entropy is plotted against system size $L$. The solid lines indicate a linear fit and the crosses indicate the predictions from the quasiparticle model.}
\label{seepeaks} }
\end{figure}

It has been pointed out in \cite{Modak:2020faf} that the lack of dips points to scrambling, which leads to loss of information about initial correlations being dispersed. It is also noted in \cite{Alba:2019ybw} that disappearance of peaks in mutual information is related to scrambling. Physically, the same should be true for logarithmic negativity. But, due to the appearance of multiple peaks (for example, Fig. \ref{massive_quench: lgnvstmassive}), it is difficult to discriminate peak height. Hence, it is more convenient to investigate the dips in entanglement entropy. More precisely, height of the dips is given by \cite{Modak:2020faf},
\begin{equation}
\delta S = \frac{S_{\ell}(\infty)-S_{\ell}(t_R)}{\ell}, \quad \text{where} \quad S_{\ell}(\infty) = \lim_{t \to \infty} \lim_{L \to \infty} S_{\ell}(t) .
\end{equation}
It is interesting to note that how fast the dip is suppressed as the ratio $\ell/L$ is increased and the parameter $\alpha$ is decreased. This is encapsulated in Fig. \ref{seepeaks}. The plot also reveals that the fall-off eventually assumes an exponential form for large $\ell/L$, which is consistent with results in \cite{Modak:2020faf}. It may be observed that the only effect of the parameter $\alpha$ seems to decrease the magnitude of the dip and there is no significant effect on the rate at which the dip disappears with increasing system size $L$.

\section{Conclusions}

In this paper, aspects of entanglement in non-local field theories have been examined. First, the investigation is carried out in a static setting, involving two subsystems with a separation between them. The main conclusions from this investigation are that for long-ranged models, the logarithmic negativity falls off more gradually with subsystem separation. Secondly, time dependence is introduced by a simple model for an instantaneous quantum quench. The variation of the negativity with respect to time is studied, with the focus being mainly on the long time behaviour to check if the negativity in long-range models showed any indication of revival. The main result of these studies may be summarised as follows: long-range models (for $0<\alpha<1$) exhibit some novel characteristics which are distinct from results obtained so far in the literature. One of the most interesting features of long-range models is that no signs of revival of entanglement negativity are found, at least in the time interval considered. 

To reinforce the results about revivals in the logarithmic negativity, the same is investigated in the context of entanglement entropy, and dips in the entanglement entropy (comparable to revivals in logarithmic negativity) are noted. The numerical results show excellent agreement with predictions from the quasiparticle picture. This establishes the feature of revivals/dips disappearing gradually in long range models. As the literature suggests that this signals scrambling, the out-of-ordered correlators (OTOC), is calculated in the appendix, which shows that fast scrambling is not present in this class of models.

This work may be extended in various directions.  For example, there also exist in the literature various other quench protocols \cite{Das:2014jna, Das:2014hqa, Das:2015jka} which have been extensively investigated from the point of view of entanglement entropy. Thus negativity might therefore also be investigated in similar settings to gain further insight into mixed state entanglement.

% \appendix
% \section{Appendix: Calculating the out-of-time-ordered correlator}

\section*{Appendix: Calculating the out-of-time-ordered correlator}

Considering the indications of scrambling for a specific regime of $\alpha$ in non-local field theories, it is desirable to further investigate the issue through calculating another quantity which has been used in the literature, namely, out-of-time-ordered correlators (OTOC) \cite{larkin1969quasiclassical}. Recently, the OTOC has seen a resurgence in the context of the AdS/CFT duality, where it has been used by Kitaev \cite{kitaev2015simple} for one-dimensional systems. It was proved that for the Sachdev-Ye-Kitaev (SYK) model \cite{Sachdev:1992fk} that the thermal OTOC exhibits exponential growth, which indicates chaos. This is because it is possible to relate the decay rate of the OTOC to the Lyapunov exponent in the semiclassical limit.

It may be noted that the growth of the OTOC has been used to probe scrambling and chaos in numerous few \cite{Rozenbaum:2016mmv, chavez2019quantum} and many-body systems \cite{Lin:2018tce, Iyoda:2017pxe}. A procedure of calculating OTOCs in quantum mechanical systems has also been formulated in \cite{Hashimoto:2017oit}. In this appendix, the same formulation is adapted for the system under consideration, i.e, a long range harmonic chain consisting of coupled oscillators. Firstly, the microcanonical (ground state) OTOC for the harmonic chain (eq. \ref{Halpha}) is defined as follows,
\begin{equation}
c_{ij}(t) = -\langle{0}| [\phi_{i}(t), \pi_j]^2 |{0}\rangle ,
\end{equation}
where the expectation value is calculated according to the ground state of the system. Inserting the resolution of unity, $\sum_m |m \rangle \langle m | = 1$ in the above equation, 
\begin{equation}
c_{ij}(t) = \sum_{m} (b_{ij}^{m})^{*}(t)b_{ij}^{m}(t), \quad b_{ij}^{m}(t) = -i \langle{0}| [\phi_{i}(t), \pi_j ]^2 | {m}\rangle .
\end{equation}
As in subsection \ref{subsec: setup}, it is convenient to make a discrete Fourier transformation to decouple the oscillators,
\begin{equation}
\phi_{i}(t) = \sum_{k=1}^{L} O_{ik} q_k(t), \quad \pi_{j} = \sum_{k=1}^{L} O_{jk} p_k .
\end{equation}
Taking into account the fact that the oscillators are decoupled in Fourier space,
\begin{equation}
b^{m}_{ij}(t) = -i \sum_{k, k^{\prime} = 1}^{L} O_{ik} O_{jk^{\prime}} \langle{0} | [q_{k}(t), p_{k^{\prime}} | {m}\rangle \delta_{k k^{\prime}}.
\end{equation}
Another insertion of the resolution of unity yields,
\begin{equation}
\langle{0} | [q_{k}(t), p_{k} ] | {m}\rangle = -\frac{i}{2} \sum_{p} q_{0p} q_{pm} \left(  E_{pm} e^{i E_{0p}t} - E_{0p} e^{i E_{pm}t} \right),
\end{equation}
where $x_{mn} = \langle{m} | x | n\rangle$, calculated from the pre-quench state of the system. Evaluating the right hand side of the above equation,
\begin{equation}
\langle{0} | [q_{k}(t), p_{k} ] | {m}\rangle = \frac{\omega_{2k}}{\omega_{1k}} \cos(\omega_{2k}t) .
\end{equation}
For the final expression for $b_{m}(t)$, the expression for the matrix $O$ needs to be inserted, which yields,
\begin{equation}
b^{0}_{ij}(t) \equiv b_{ij}(t) = -i\sum_{k=1}^{L} \frac{\omega_{2k}}{\omega_{1k}} \cos(\omega_{2k}t) \cos \left[  \frac{2\pi(i-j)k}{L} \right]
\end{equation}
From the above expression, it is easy to obtain the OTOC for the harmonic chain.

\begin{figure}[h]{
		\begin{tabular}{cc}
			\includegraphics[width=0.5\textwidth]{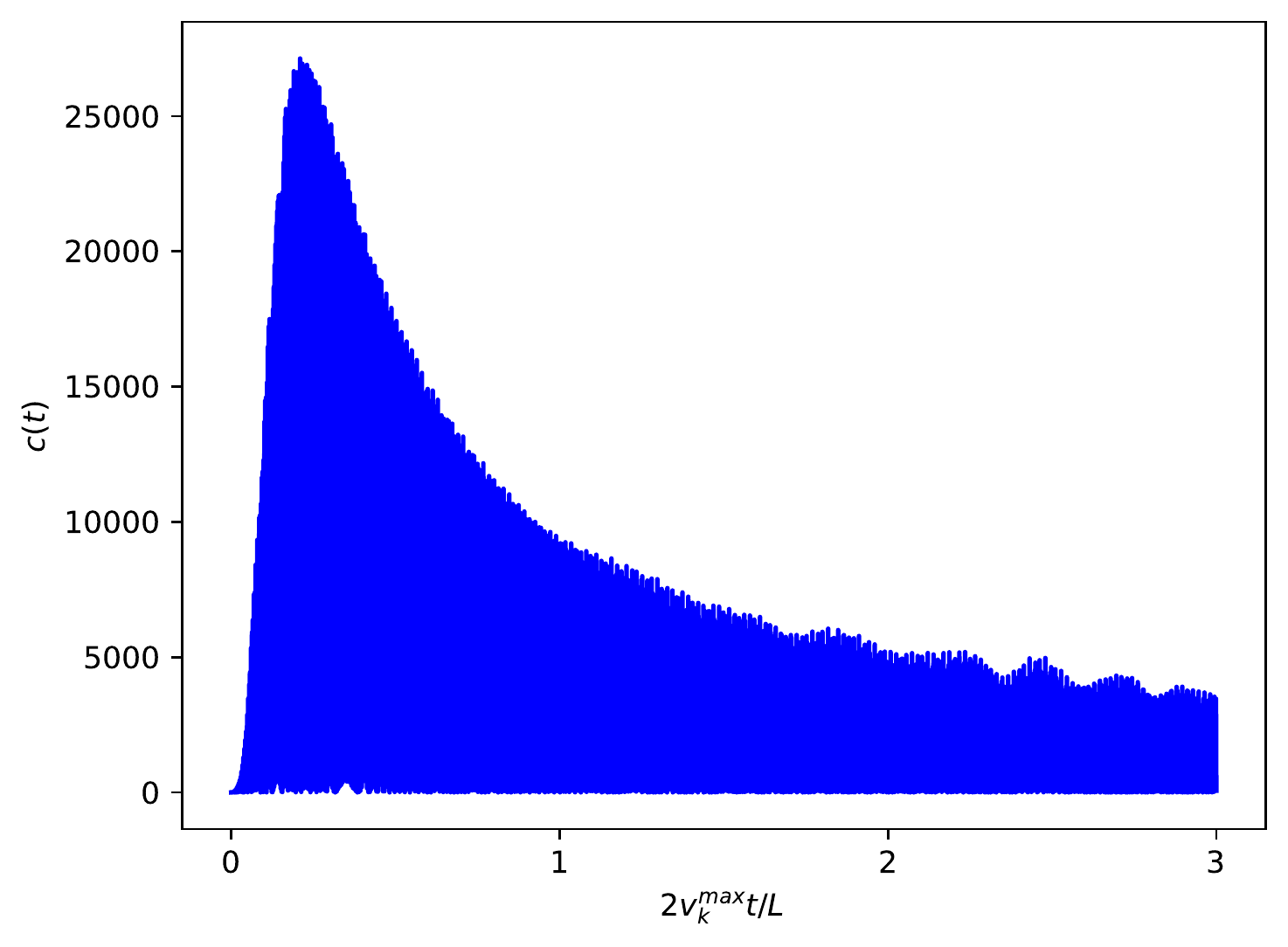}&
			\includegraphics[width=0.5\textwidth]{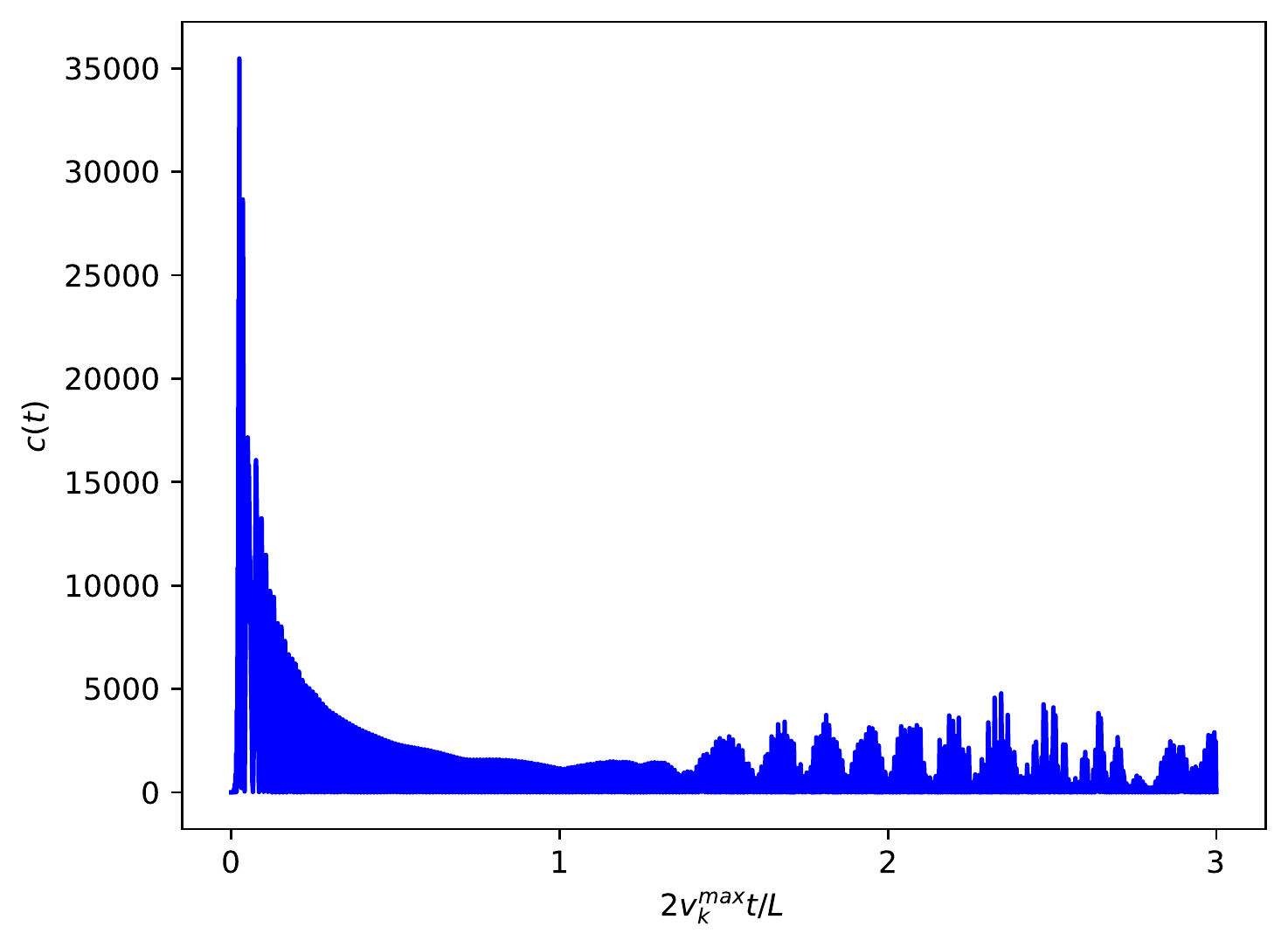}\\
			(a) & (b) \\
			\includegraphics[width=0.5\textwidth]{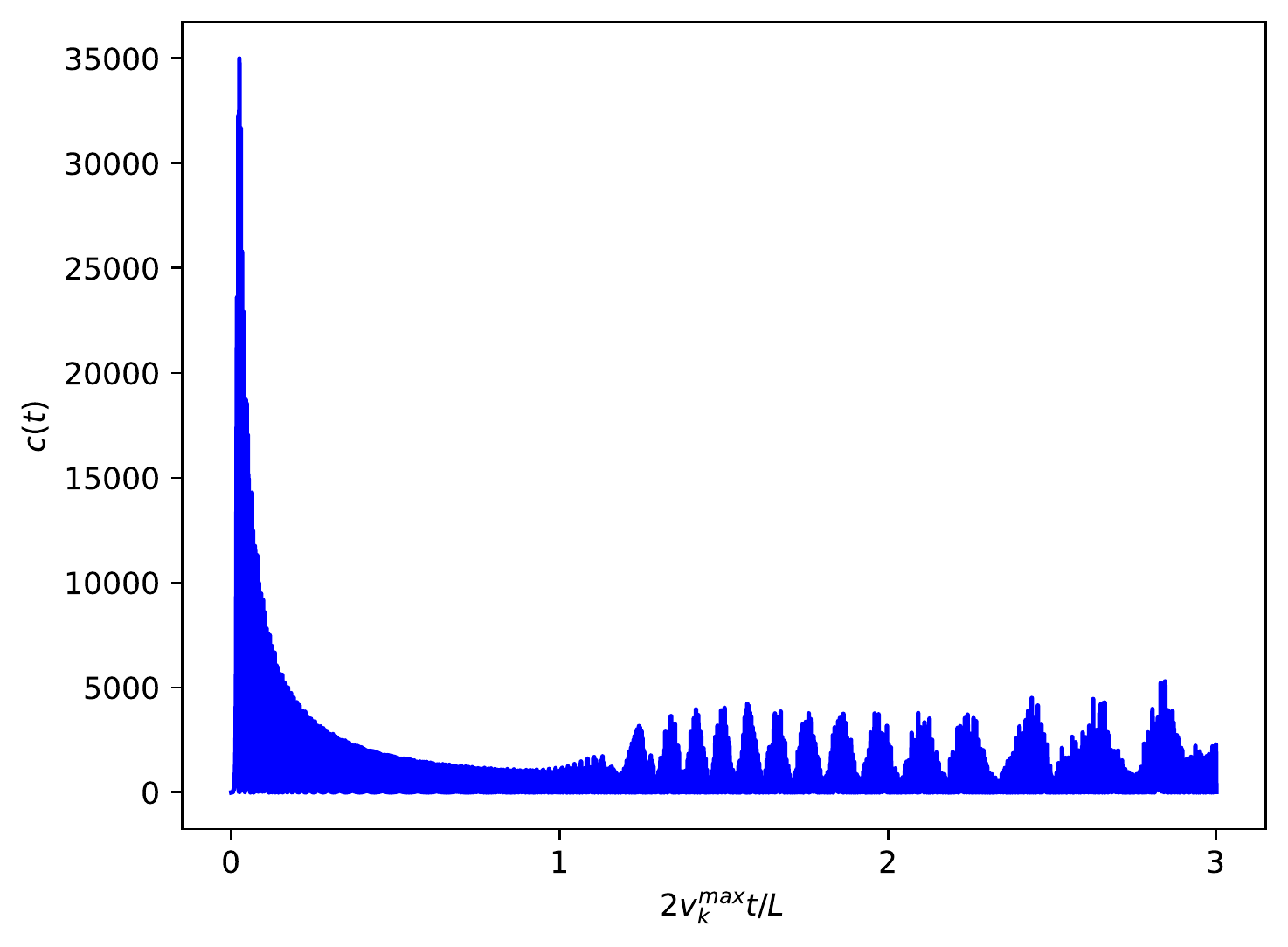} &
			\includegraphics[width=0.5\textwidth]{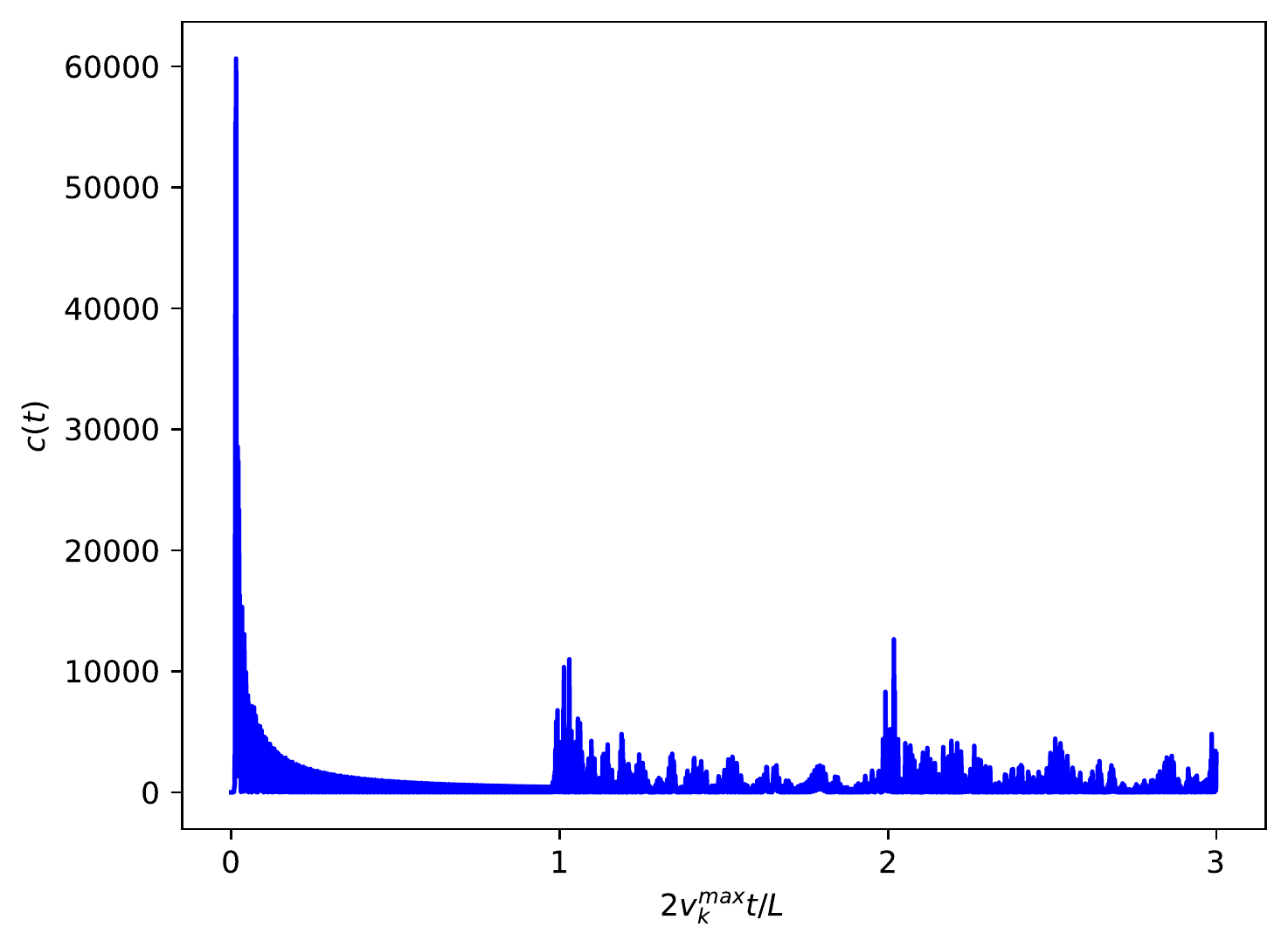} \\
			(c) & (d)
		\end{tabular}
		\caption{$c(t) \equiv c_{ij}(t)$ vs $t$ for different values of the $\alpha$ and $n=2000$ and $|i-j|=25$. For (a), (b), (c) and (d), $\alpha$ takes values $0.6$, $0.9$, $0.95$ and $1.1$ respectively. The mass parameter is quenched from $m=4$ to $m=2$.}
		\label{fig: otoc}}
\end{figure}

With the above expressions in hand, it is possible to plot the the microcanonical OTOC against time, as shown in Fig. \ref{fig: otoc}. It may be observed that the microcanonical OTOC shows oscillates extremely rapidly in a small time interval. Nevertheless, the general feature appears to be that the OTOC shows ``revivals'' at $t = 2 v_k^{max}/L$ as $\alpha>1.0$, similar to the case for logarithmic negativity and entanglement entropy. It is also clear that there seems to be no exponential growth of the OTOC, indicating the lack of fast scrambling. This is physically not surprising, since it is usually the case that fast scrambling is present in non-integrable models. 

% \bibliography{ref}

\begin{thebibliography}{10}

\bibitem{Hastings:2005pr}
M.~B. Hastings and T.~Koma, \emph{{Spectral gap and exponential decay of
  correlations}},
  \href{http://dx.doi.org/10.1007/s00220-006-0030-4}{\emph{Commun. Math. Phys.}
  {\bf 265} (2006) 781--804}, [\href{http://arxiv.org/abs/math-ph/0507008}{{\tt
  math-ph/0507008}}].

\bibitem{Horodecki}
R.~Horodecki, P.~Horodecki, M.~Horodecki and K.~Horodecki, \emph{Quantum
  entanglement}, \href{http://dx.doi.org/10.1103/RevModPhys.81.865}{\emph{Rev.
  Mod. Phys.} {\bf 81} (Jun, 2009) 865--942}.

\bibitem{Plenionegativity}
M.~B. Plenio, \emph{Logarithmic negativity: A full entanglement monotone that
  is not convex},
  \href{http://dx.doi.org/10.1103/PhysRevLett.95.090503}{\emph{Phys. Rev.
  Lett.} {\bf 95} (Aug, 2005) 090503}.

\bibitem{Audenaert:2002xfl}
K.~Audenaert, J.~Eisert, M.~Plenio and R.~Werner, \emph{{Entanglement
  Properties of the Harmonic Chain}},
  \href{http://dx.doi.org/10.1103/PhysRevA.66.042327}{\emph{Phys. Rev. A} {\bf
  66} (2002) 042327}, [\href{http://arxiv.org/abs/quant-ph/0205025}{{\tt
  quant-ph/0205025}}].

\bibitem{Angel-Ramelli:2020wfo}
J.~Angel-Ramelli, C.~Berthiere, V.~G.~M. Puletti and L.~Thorlacius,
  \emph{{Logarithmic Negativity in Quantum Lifshitz Theories}},
  \href{http://dx.doi.org/10.1007/JHEP09(2020)011}{\emph{JHEP} {\bf 09} (2020)
  011}, [\href{http://arxiv.org/abs/2002.05713}{{\tt 2002.05713}}].

\bibitem{MohammadiMozaffar:2017chk}
M.~R. Mohammadi~Mozaffar and A.~Mollabashi, \emph{{Logarithmic Negativity in
  Lifshitz Harmonic Models}},
  \href{http://dx.doi.org/10.1088/1742-5468/aac135}{\emph{J. Stat. Mech.} {\bf
  1805} (2018) 053113}, [\href{http://arxiv.org/abs/1712.03731}{{\tt
  1712.03731}}].

\bibitem{Marcovitch:2008sxc}
S.~Marcovitch, A.~Retzker, M.~Plenio and B.~Reznik, \emph{{Critical and
  noncritical long-range entanglement in Klein-Gordon fields}},
  \href{http://dx.doi.org/10.1103/PhysRevA.80.012325}{\emph{Phys. Rev. A} {\bf
  80} (2009) 012325}, [\href{http://arxiv.org/abs/0811.1288}{{\tt 0811.1288}}].

\bibitem{Calabrese:2012ew}
P.~Calabrese, J.~Cardy and E.~Tonni, \emph{{Entanglement negativity in quantum
  field theory}},
  \href{http://dx.doi.org/10.1103/PhysRevLett.109.130502}{\emph{Phys. Rev.
  Lett.} {\bf 109} (2012) 130502}, [\href{http://arxiv.org/abs/1206.3092}{{\tt
  1206.3092}}].

\bibitem{Calabrese:2012nk}
P.~Calabrese, J.~Cardy and E.~Tonni, \emph{{Entanglement negativity in extended
  systems: A field theoretical approach}},
  \href{http://dx.doi.org/10.1088/1742-5468/2013/02/P02008}{\emph{J. Stat.
  Mech.} {\bf 1302} (2013) P02008}, [\href{http://arxiv.org/abs/1210.5359}{{\tt
  1210.5359}}].

\bibitem{Calabrese:2014yza}
P.~Calabrese, J.~Cardy and E.~Tonni, \emph{{Finite temperature entanglement
  negativity in conformal field theory}},
  \href{http://dx.doi.org/10.1088/1751-8113/48/1/015006}{\emph{J. Phys. A} {\bf
  48} (2015) 015006}, [\href{http://arxiv.org/abs/1408.3043}{{\tt 1408.3043}}].

\bibitem{Shiba:2013jja}
N.~Shiba and T.~Takayanagi, \emph{{Volume Law for the Entanglement Entropy in
  Non-local QFTs}},
  \href{http://dx.doi.org/10.1007/JHEP02(2014)033}{\emph{JHEP} {\bf 02} (2014)
  033}, [\href{http://arxiv.org/abs/1311.1643}{{\tt 1311.1643}}].

\bibitem{Basa:2019ywr}
B.~Basa, G.~La~Nave and P.~W. Phillips, \emph{{Classification of nonlocal
  actions: Area versus volume entanglement entropy}},
  \href{http://dx.doi.org/10.1103/PhysRevD.101.106006}{\emph{Phys. Rev. D} {\bf
  101} (2020) 106006}, [\href{http://arxiv.org/abs/1907.09494}{{\tt
  1907.09494}}].

\bibitem{Lu:2019xwg}
T.-C. Lu and T.~Grover, \emph{{Structure of Quantum Entanglement at a Finite
  Temperature Critical Point}},  \href{http://arxiv.org/abs/1907.01569}{{\tt
  1907.01569}}.

\bibitem{Caffarelli_2007}
L.~Caffarelli and L.~Silvestre, \emph{An extension problem related to the
  fractional laplacian},
  \href{http://dx.doi.org/10.1080/03605300600987306}{\emph{Communications in
  Partial Differential Equations} {\bf 32} (Aug, 2007) 1245–1260}.

\bibitem{Nezhadhaghighi:2014pwa}
M.~Ghasemi~Nezhadhaghighi and M.~Rajabpour, \emph{{Entanglement dynamics in
  short and long-range harmonic oscillators}},
  \href{http://dx.doi.org/10.1103/PhysRevB.90.205438}{\emph{Phys. Rev. B} {\bf
  90} (2014) 205438}, [\href{http://arxiv.org/abs/1408.3744}{{\tt 1408.3744}}].

\bibitem{PhysRevLett.103.120502}
K.~Kim, M.-S. Chang, R.~Islam, S.~Korenblit, L.-M. Duan and C.~Monroe,
  \emph{Entanglement and tunable spin-spin couplings between trapped ions using
  multiple transverse modes},
  \href{http://dx.doi.org/10.1103/PhysRevLett.103.120502}{\emph{Phys. Rev.
  Lett.} {\bf 103} (Sep, 2009) 120502}.

\bibitem{PhysRevLett.109.267203}
T.~Koffel, M.~Lewenstein and L.~Tagliacozzo, \emph{Entanglement entropy for the
  long-range ising chain in a transverse field},
  \href{http://dx.doi.org/10.1103/PhysRevLett.109.267203}{\emph{Phys. Rev.
  Lett.} {\bf 109} (Dec, 2012) 267203}.

\bibitem{PhysRevA.89.062330}
R.~G. Unanyan and M.~Fleischhauer, \emph{Entanglement dynamics in
  harmonic-oscillator chains},
  \href{http://dx.doi.org/10.1103/PhysRevA.89.062330}{\emph{Phys. Rev. A} {\bf
  89} (Jun, 2014) 062330}.

\bibitem{PhysRevLett.113.210402}
N.~Nessi, A.~Iucci and M.~A. Cazalilla, \emph{Quantum quench and
  prethermalization dynamics in a two-dimensional fermi gas with long-range
  interactions},
  \href{http://dx.doi.org/10.1103/PhysRevLett.113.210402}{\emph{Phys. Rev.
  Lett.} {\bf 113} (Nov, 2014) 210402}.
  
%\cite{MohammadiMozaffar:2018vmk}
\bibitem{MohammadiMozaffar:2018vmk}
M.~R.~Mohammadi Mozaffar and A.~Mollabashi,
%``Entanglement Evolution in Lifshitz-type Scalar Theories,''
JHEP \textbf{01}, 137 (2019)
doi:10.1007/JHEP01(2019)137
[arXiv:1811.11470 [hep-th]].
%23 citations counted in INSPIRE as of 29 Jan 2022

%\cite{Mozaffar:2021nex}
\bibitem{Mozaffar:2021nex}
M.~R.~M.~Mozaffar and A.~Mollabashi,
%``On the Time Scaling of Entanglement in Integrable Scale-Invariant Theories,''
[arXiv:2106.14700 [hep-th]].
%2 citations counted in INSPIRE as of 29 Jan 2022

\bibitem{nachtergaele2006propagation}
B.~Nachtergaele, Y.~Ogata and R.~Sims, \emph{Propagation of correlations in
  quantum lattice systems}, {\emph{Journal of statistical physics} {\bf 124}
  (2006) 1--13}.

\bibitem{foss2015nearly}
M.~Foss-Feig, Z.-X. Gong, C.~W. Clark and A.~V. Gorshkov, \emph{Nearly linear
  light cones in long-range interacting quantum systems}, {\emph{Physical
  review letters} {\bf 114} (2015) 157201}.

\bibitem{gong2014persistence}
Z.-X. Gong, M.~Foss-Feig, S.~Michalakis and A.~V. Gorshkov, \emph{Persistence
  of locality in systems with power-law interactions}, {\emph{Physical review
  letters} {\bf 113} (2014) 030602}.

\bibitem{guo2020signaling}
A.~Y. Guo, M.~C. Tran, A.~M. Childs, A.~V. Gorshkov and Z.-X. Gong,
  \emph{Signaling and scrambling with strongly long-range interactions},
  {\emph{Physical Review A} {\bf 102} (2020) 010401}.
  
%\cite{Frassino:2019yip}
\bibitem{Frassino:2019yip}
A.~M.~Frassino and O.~Panella,
%``Quantization of nonlocal fractional field theories via the extension problem,''
Phys. Rev. D \textbf{100}, no.11, 116008 (2019)
doi:10.1103/PhysRevD.100.116008
[arXiv:1907.00733 [hep-th]].
%5 citations counted in INSPIRE as of 29 Jan 2022

\bibitem{serafini2017quantum}
A.~Serafini, \emph{Quantum continuous variables: a primer of theoretical
  methods}.
\newblock CRC press, 2017.

\bibitem{Wichterich:2008vfx}
H.~Wichterich, J.~Molina-Vilaplana and S.~Bose, \emph{{Scaling of entanglement
  between separated blocks in spin chains at criticality}},
  \href{http://dx.doi.org/10.1103/PhysRevA.80.010304}{\emph{Phys. Rev. A} {\bf
  80} (2009) 010304}, [\href{http://arxiv.org/abs/0811.1285}{{\tt 0811.1285}}].

\bibitem{Calabrese:2005in}
P.~Calabrese and J.~L. Cardy, \emph{{Evolution of entanglement entropy in
  one-dimensional systems}},
  \href{http://dx.doi.org/10.1088/1742-5468/2005/04/P04010}{\emph{J. Stat.
  Mech.} {\bf 0504} (2005) P04010},
  [\href{http://arxiv.org/abs/cond-mat/0503393}{{\tt cond-mat/0503393}}].

\bibitem{Fagotti:2010yr}
M.~Fagotti and P.~Calabrese, \emph{{Entanglement entropy of two disjoint blocks
  in XY chains}},
  \href{http://dx.doi.org/10.1088/1742-5468/2010/04/P04016}{\emph{J. Stat.
  Mech.} {\bf 1004} (2010) P04016}, [\href{http://arxiv.org/abs/1003.1110}{{\tt
  1003.1110}}].

\bibitem{Modak:2020faf}
R.~Modak, V.~Alba and P.~Calabrese, \emph{{Entanglement revivals as a probe of
  scrambling in finite quantum systems}},
  \href{http://dx.doi.org/10.1088/1742-5468/aba9d9}{\emph{J. Stat. Mech.} {\bf
  2008} (2020) 083110}, [\href{http://arxiv.org/abs/2004.08706}{{\tt
  2004.08706}}].

\bibitem{Coser:2014gsa}
A.~Coser, E.~Tonni and P.~Calabrese, \emph{{Entanglement negativity after a
  global quantum quench}},
  \href{http://dx.doi.org/10.1088/1742-5468/2014/12/P12017}{\emph{J. Stat.
  Mech.} {\bf 1412} (2014) P12017}, [\href{http://arxiv.org/abs/1410.0900}{{\tt
  1410.0900}}].

\bibitem{PhysRevB.91.045131}
M.~A. Rajabpour and S.~Sotiriadis, \emph{Quantum quench in long-range field
  theories}, \href{http://dx.doi.org/10.1103/PhysRevB.91.045131}{\emph{Phys.
  Rev. B} {\bf 91} (Jan, 2015) 045131}.

\bibitem{Alba:2019ybw}
V.~Alba and P.~Calabrese, \emph{{Quantum information scrambling after a quantum
  quench}}, \href{http://dx.doi.org/10.1103/PhysRevB.100.115150}{\emph{Phys.
  Rev. B} {\bf 100} (2019) 115150},
  [\href{http://arxiv.org/abs/1903.09176}{{\tt 1903.09176}}].

\bibitem{Das:2014jna}
S.~R. Das, D.~A. Galante and R.~C. Myers, \emph{{Universal scaling in fast
  quantum quenches in conformal field theories}},
  \href{http://dx.doi.org/10.1103/PhysRevLett.112.171601}{\emph{Phys. Rev.
  Lett.} {\bf 112} (2014) 171601}, [\href{http://arxiv.org/abs/1401.0560}{{\tt
  1401.0560}}].

\bibitem{Das:2014hqa}
S.~R. Das, D.~A. Galante and R.~C. Myers, \emph{{Universality in fast quantum
  quenches}}, \href{http://dx.doi.org/10.1007/JHEP02(2015)167}{\emph{JHEP} {\bf
  02} (2015) 167}, [\href{http://arxiv.org/abs/1411.7710}{{\tt 1411.7710}}].

\bibitem{Das:2015jka}
S.~R. Das, D.~A. Galante and R.~C. Myers, \emph{{Smooth and fast versus
  instantaneous quenches in quantum field theory}},
  \href{http://dx.doi.org/10.1007/JHEP08(2015)073}{\emph{JHEP} {\bf 08} (2015)
  073}, [\href{http://arxiv.org/abs/1505.05224}{{\tt 1505.05224}}].

\bibitem{larkin1969quasiclassical}
A.~Larkin and Y.~N. Ovchinnikov, \emph{Quasiclassical method in the theory of
  superconductivity}, {\emph{Sov Phys JETP} {\bf 28} (1969) 1200--1205}.

\bibitem{kitaev2015simple}
A.~Kitaev, \emph{A simple model of quantum holography},  in \emph{KITP strings
  seminar and Entanglement}, vol.~12, p.~26, 2015.

\bibitem{Sachdev:1992fk}
S.~Sachdev and J.~Ye, \emph{{Gapless spin fluid ground state in a random,
  quantum Heisenberg magnet}},
  \href{http://dx.doi.org/10.1103/PhysRevLett.70.3339}{\emph{Phys. Rev. Lett.}
  {\bf 70} (1993) 3339}, [\href{http://arxiv.org/abs/cond-mat/9212030}{{\tt
  cond-mat/9212030}}].

\bibitem{Rozenbaum:2016mmv}
E.~B. Rozenbaum, S.~Ganeshan and V.~Galitski, \emph{{Lyapunov Exponent and
  Out-of-Time-Ordered Correlator\textquoteright{}s Growth Rate in a Chaotic
  System}}, \href{http://dx.doi.org/10.1103/PhysRevLett.118.086801}{\emph{Phys.
  Rev. Lett.} {\bf 118} (2017) 086801},
  [\href{http://arxiv.org/abs/1609.01707}{{\tt 1609.01707}}].

\bibitem{chavez2019quantum}
J.~Ch{\'a}vez-Carlos, B.~L{\'o}pez-del Carpio, M.~A. Bastarrachea-Magnani,
  P.~Str{\'a}nsk{\`y}, S.~Lerma-Hern{\'a}ndez, L.~F. Santos et~al.,
  \emph{Quantum and classical lyapunov exponents in atom-field interaction
  systems}, {\emph{Physical review letters} {\bf 122} (2019) 024101}.

\bibitem{Lin:2018tce}
C.-J. Lin and O.~I. Motrunich, \emph{{Out-of-time-ordered correlators in a
  quantum Ising chain}},
  \href{http://dx.doi.org/10.1103/PhysRevB.97.144304}{\emph{Phys. Rev. B} {\bf
  97} (2018) 144304}, [\href{http://arxiv.org/abs/1801.01636}{{\tt
  1801.01636}}].

\bibitem{Iyoda:2017pxe}
E.~Iyoda and T.~Sagawa, \emph{{Scrambling of Quantum Information in Quantum
  Many-Body Systems}},
  \href{http://dx.doi.org/10.1103/PhysRevA.97.042330}{\emph{Phys. Rev. A} {\bf
  97} (2018) 042330}, [\href{http://arxiv.org/abs/1704.04850}{{\tt
  1704.04850}}].

\bibitem{Hashimoto:2017oit}
K.~Hashimoto, K.~Murata and R.~Yoshii, \emph{{Out-of-time-order correlators in
  quantum mechanics}},
  \href{http://dx.doi.org/10.1007/JHEP10(2017)138}{\emph{JHEP} {\bf 10} (2017)
  138}, [\href{http://arxiv.org/abs/1703.09435}{{\tt 1703.09435}}].

\end{thebibliography}
% \bibliographystyle{JHEP}
\providecommand{\noopsort}[1]{}\providecommand{\singleletter}[1]{#1}%
\providecommand{\href}[2]{#2}\begingroup\raggedright\endgroup

\end{document}